# Kelvin-Helmholtz instability in binary fluids with miscibility gap

Anubhav Dubey[1] and Sakir Amiroudine[1]

*Univ. Bordeaux, CNRS, Bordeaux INP, I2M, UMR 5295, F-33400, Talence, France*

(*Electronic mail: anubhav.dubey@u-bordeaux.fr)

The isothermal spatio-temporal evolution of an interface between binary fluids, with temperature sensitive miscibility gap, subjected to shear flow is investigated using direct numerical simulations. The thermophysical properties and the interfacial tension in such fluids exhibit a dynamic temperature dependence based on the degree of miscibility. These fluid pairs can therefore be deployed in crucial microfluidic applications involving enhancement of liquid-liquid mass transfer. In this study, a modified phase-field approach that allows modeling of partially miscible fluids and continuous transition from an initially immiscible state to a completely miscible state is employed to investigate the Kelvin-Helmholtz (KH) instability. The analysis entails two different configurations based on the thermodynamic equilibrium of the system. In the first configuration, the fluid pair is considered in equilibrium at different degrees of partially miscibility. The second configuration deals with fluids out of equilibrium thereby allowing simultaneous mass transport across the interface. The results reveal the effect of apparent surface tension in miscible fluids, shear flow profile, the degree of stratification and the effect of mass diffusion across the interface. The shear instability in a fluid pair close to the consolute point is found to be independent of the extent of stratification. Finally, the competition between the time scales of the instability and the diffusion process reveals the importance of the surface tension of binary fluids in the immiscible limit.

## I. INTRODUCTION

The shearing instability underpins the mixing process (or the homogenization process) in several industrial applications and natural phenomena. Traditionally, such instabilities have been investigated within the purview of either immiscible fluid pairs or completely miscible fluid pairs. However, the recent discoveries of the potential applications of binary fluids exhibiting a miscibility gap has renewed the scientific attention. For instance, binary fluids with temperature sensitive miscibility gap may be employed for extraction and separation of bio-active compounds[1,2], liquid-crystal microdroplet formation[3] and enhancement of liquid-liquid mass transfer[4]. Similarly, a thermo-polymer exhibiting a pH sensitive miscibility suits targeted drug-delivery applications[5]. The relative degree of miscibility between such fluid pairs is therefore a function of temperature (or pH). Thus, the fluid pairs can be categorized on the basis of the directional evolution of the relative degree of miscibility as either the lower critical solution temperature (LCST) fluids or the upper critical solution temperature (UCST) fluids. The UCST is the maximum temperature till which a thermodynamically stable interface segregates the fluid-1 dominated region in space from the fluid-2 dominated region. If the temperature is raised above the UCST, the mass transfer across the interface continues till the concentration gradient is trivial. Therefore, the Korteweg stress[6] (which is a function of the concentration gradient) in both, pre-UCST and post-UCST regime is a function of the temperature. Consequently, such fluid pairs also find applications in microfluidic flow pattern tuning with temperature being the tuning parameter[7]. However, despite the numerous applications, binary fluids with temperature sensitive miscibility gap are relatively under-explored specifically in the context of shearing instabilities. In the current study, we investigate the Kelvin-Helmholtz (KH) instability to explore the dynamics and the spatio-temporal evolution of the flow field for a temperature sensitive interface, to understand the nuances of the interplay between the shear and the relative degree of miscibility. Before venturing into further details, we present a brief overview of the rich literature of the shearing instabilities in both immiscible fluid pairs and the miscible fluid pairs.

The shearing instability is associated with the formation of the structures that draws energy from the base flow, which is subsequently lost due to irreversible mixing[8]. They act as a route of transition from the laminar to the turbulent regime in density-stratified flows. The Kelvin-Helmholtz (KH) instability is the most widely explored shear instability. Taylor[9] and Goldstein[10] were among the first ones to linearly analyze the effect of the imposition of a continuous velocity profile over stably density-stratified inviscid fluids. The analysis revealed the existence of a critical Richardson number, which is the ratio of the destabilizing agents (velocity shear) to the stabilizing agents (density stratification), for the propensity of the system to be unstable. The threshold Richardson number was found to be $\frac{1}{4}$. Drazin[11] considered smoothly varying velocity profile and found the critical Richardson number to be the same. Several studies[12,13] were performed to tackle the meteorological and oceanographic problems and therefore the density variation effects were only considered in the buoyancy terms while being neglected in the inertial terms. Thorpe[14] experimentally observed the distinct phases of the growth of the disturbance. The initially sinusoidal waves distort into a spiral form (billow) while maintaining symmetry about the original position of the interface. The development of the spiral shape is accompanied by the accumulation of the other fluid on the side. A considerable sharpening of the interface is observed as the process continues. The development of the billow is followed by the emanation of the secondary instabilities which precedes the transition to turbulent mixing. Maslowe and Kelly[15] extended the analysis by considering density variations in the inertial terms as well. The limiting cases were categorized on the basis of the Froude number. The low Froude number cases correspond to buoyancy dominated regimes whereas the high Froude number correspond to inertia dominated regime. The study revealed that the flow can be destabilized if the lighter fluid has the higher velocity. Therefore, the Froude number can play both a destabilizing as well as the stabilizing role depending upon the velocity and density profiles at the initial state.

Maslowe and Thompson[16] further relaxed the assumption of inviscid fluids to account for the effect of viscosity of the fluids. The damping effect due to viscosity was found to be relatively weaker in a stratified shear layer. Furthermore, the viscosity could only effect the small amplitude waves that are close to the inviscid stability boundary. Thus, the competition between the destabilizing shearing effect of velocity and the stabilizing effect of density gradient dictates the onset of the instability. The numerical investigation of inviscid fluids by Hazel[17] revealed the effect of boundaries on the stability characteristics. As the depth of the fluid layer decreases, due to close proximity of the initially stable interface with the boundary walls, the larger wavelength disturbances are destabilized while the shorter wavelength disturbances are stabilized. Naturally, most of studies considered equal depth of the two layers, whether the stratified fluids be bounded or unbounded, leading to a symmetric stratified shear layer flow. Based on the interaction between the density profile and the velocity profile, one may either observe a stationary instability with a fixed vortex or an oscillatory instability with a traveling vortex. While the stationary instability is the KH instability, the second kind is commonly referred to as the Holmboe instability[18]. Smith and Peltier[19] analyzed the transition criteria from the KH instability to the Holmboe instability. The analysis revealed that the KH instability is typically observed in weakly-stratified flows whereas the Holmboe instability waves dominate for sufficiently strong stratification only. Further, the transition is dependent on the relative length scales of the velocity and the density profiles. Baines and Mitsudera[20] and Caulfield[21] argued phase-locking as the mechanism responsible to ensure the KH instability. In the case of relatively higher Richardson number, or larger velocity length scale, the phase-locking is not possible leading to Holmboe instability. Hogg and Ivey[8] found that the growth rate for Holmboe instability is generally smaller than that of the KH instability. Furthermore, their analysis also revealed that there is no critical value of the Richardson number for the Holmboe instability under inviscid assumption. The inclusion of viscosity and diffusive effects, on the other hand, strongly suppress the region of instability for Holmboe modes[22]. A comparison of growth rate between the KH instability and Holmboe instability reveals that the KH instability typically results in relatively faster transition to turbulence. Further the growth rate of the instability is also dependent on the viscosity-disparity between the two layers. The most stable state a shear layer flow can attain corresponds to both fluids having the same viscosity[23]. The transition between KH and Holmboe instabilities has further been investigated through non-linear means[24] by describing the shear mechanism as an interaction between propagative waves. Barrows and Choi[25] extended the analysis of Holmboe waves by relaxing the Boussinesq approximation. Finally, the role of surface tension has been explicated by Chandrasekhar[26] for various flow and density profiles thereby establishing the fundamental flow behavior in the context of shear at the immiscible interfaces.

On the other hand, the interface between two miscible fluids subjected to shear exhibits peculiar characteristics. Such processes are relevant in the chemical industries[27]. Moatimid and El-dib[27] conducted a linear analysis on inviscid fluids and revealed the destabilizing effect of heat and mass transport across the interface. Harang et al.[28] relaxed the inviscid assumption and performed numerical simulations to model the behavior of mudflow at the bottom of estuaries. The viscosity is quantified by defining the Reynolds number for the flow. At intermediate Reynolds number, the viscosity stratification strongly effects the onset of the instability. However, at relatively larger Reynolds number, the influence of viscosity stratification is less pronounced. Further, the viscosity of the fluid also impacts the evolution of the thickness of the interface. Sahu and Govindarajan[29] found the viscosity stratification to play the dual role, i.e. stabilizing or destabilizing, depending upon the location of the velocity profile inflection point within the higher or the lower viscous fluids. Surprisingly, the mass diffusion across the interface was found to have marginal impact on the stability results. Recently, Caulfield[30] reviewed the mixing in stratified flows and the subsequent turbulence regime. The variance of the density field is proposed to be a suitable quantification of the degree of mixing. However, to the best of our knowledge, most studies neglect the apparent interfacial tension[31-34] in the miscible regime. The incorporation of this interfacial tension allows to preserve the definition of the interface as long as the concentration gradient is non-trivial and the flow has not ventured into the turbulent regime. Further, the rate of mass transport is also affected due to the presence of the interface as the classical Fick's law of diffusion is no longer applicable[35,36]. Such effects could lead to a significant shift in the stability boundaries, specifically in the context of binary fluids with temperature sensitive miscibility gap.

Thus, in the current study, we investigate the behavior of the temperature sensitive liquid-liquid interface subjected to a symmetric shear. A symmetric shear is chosen due to the associated simplicity in demarcating the KH instability from the Holmboe instability. The temperature sensitivity of the interface leads to the possibility of simultaneous mass transport across the interface, should a thermal stratification be applied. Recently, such binary fluids have been explored in the context of Rayleigh-Taylor instability[37] and the KH instability[36,38] employing a modified phase-field approach[35]. However, these studies assume an arbitrary initial profile for the immiscible state and does not account for the initial temperature of the system. Further, the model proposed by Vorobev[35] employs qualitative empirical constants. Bestehorn et al.[39] proposed further modification to the phase-field approach to account for the proximity of the system temperature to the UCST of the fluid pair. Unfortunately, the model fails to account for the experimentally observed effective interfacial tension which is transient in nature[31–34]. Borcia *et al.*[40] proposed another model, similar to the model of Bestehorn *et al.*[39]. However, the second model assumes an incorrect temperature dependence resulting in failure to model the partially miscible states. Thus, in order to mitigate the limitations of the previous models, we proposed a comprehensive phase-field model[41,42] which on one hand correctly model the partially miscible states and also tracks the evolution from an initially immiscible state to a miscible state while accounting for the effective interfacial tension through Korteweg stress formulation. We employ the modified phase-field approach to parametrically analyze the non-linear KH instability. The model is implemented in the open-source PhaseFieldFoam[43] code (OpenFOAM framework) to perform the two-dimensional numerical simulations. The nuances pertaining to the model and the subsequent solver details are presented in the ensuing sections.

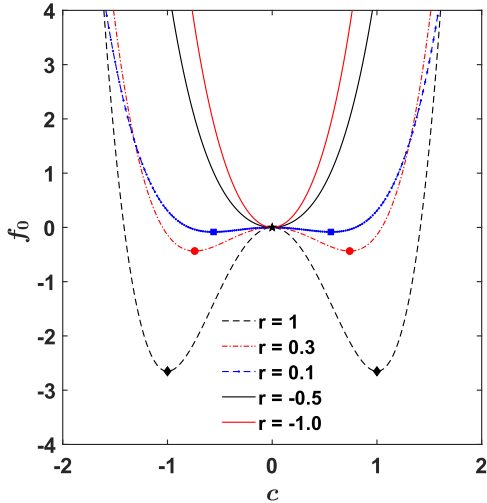

FIG. 1. Bulk free energy density

The rest of the paper is organized as follows: Sec. II describes the modified phase-field model which is employed to reformulate the governing equations in Sec. III. Sec. IV elucidates the numerical methodology employed to obtain the results discussed in Sec. V. In Sec. VI concluding remarks are provided.

## II. PHASE-FIELD MODELING: A MODIFIED APPROACH

The interface between the two bulk fluid regions evolve with the hydrodynamic flow. We employ phase-field method[44,45] to capture the spatio-temporal evolution of the interface. In this method, the interface is modeled as a thin transition zone with finite thickness $\tilde{\epsilon}$. An order parameter, $c \in [-1, 1]$, is defined to accommodate the changes in the intensive variables due to the associated homogeneity thereby allowing us to treat the two-fluids problem with one fluid formulation. The thermophysical properties vary rapidly but smoothly over the interfacial region. The order parameter $c$ is employed to define a total free energy functional entailing the contributions of the bulk free energies of the two fluids and the excess energy associated with the presence of the interface (i.e. the surface separating the two bulk regions). Herein, all the dimensional variables are accented by a tilde ($\sim$) sign. The total free energy functional can be expressed as[46–48]:

$$\tilde{F}(c, \boldsymbol{\nabla} c) = \int_{\tilde{\Omega}} \left[ \tilde{f}_o(c) + \frac{\tilde{\Lambda}}{2} |\tilde{\boldsymbol{\nabla}} c|^2 \right] d\tilde{\Omega}, \tag{1}$$

where $\tilde{f}_0(c)$ is the bulk free energy density and $\tilde{\Lambda}$ is the mixing energy density. The qualitative profile of the bulk free energy density as a function of the order parameter determines the thermodynamically preferred state of the system of fluids as shown in fig. 1. Ginzburg and Landau[49] proposed a formulation resulting in double-well potential structure of the bulk free energy density. A double well potential function leads to phase-segregation into states corresponding to minimum bulk free energy following the second law of thermodynamics. Vorobev[35] exploited this feature to modify $\tilde{f}_0(c)$ by incorporating empirical constants to transform the double-well function into a single well potential function to model binary fluids with UCST. In our previous studies[41,42], we proposed a modified formulation of $\tilde{f}_0(c)$ to passively link the binary fluids description with the system temperature. The modified bulk free energy density can be expressed as:

$$\tilde{f}_0(c) = \frac{\tilde{\Lambda}}{\tilde{\epsilon}^2} (\frac{1}{4} |r|^a c^4 - \frac{1}{2} r c^2), \tag{2}$$

where "$a$" ($0 < a < 1$) is an empirical constant and $r$ is the dimensionless miscibility parameter governing the proximity of the system temperature to the UCST ($\tilde{T}_c$). The miscibility parameter $r$ is a monotonic function of the system temperature and can be expressed as:

$$r = Z(\theta), \tag{3}$$

where $\theta$ is the reduced temperature given by $\theta = \frac{\tilde{T} - \tilde{T}_c}{\tilde{T}_c}$. The function $Z$ must be defined in such a way that $r \geq 0$ for $\tilde{T} \leq \tilde{T}_c$ and $r < 0$ for $\tilde{T} > \tilde{T}_c$. Such dependence of the miscibility parameter $r$ on the system temperature $\tilde{T}$ would inherently lead to the transformation of the bulk free energy density from the initial double well potential function to single well potential function. Thus, it allows one to model the continuous transition from an immiscible/partially miscible state to miscible state as the system temperature varies from $\tilde{T} < \tilde{T}_c$ to $\tilde{T} > \tilde{T}_c$. A possible variant of the function $Z$ is[42]:

$$r = Z(\theta) = \frac{e^{(-c\theta)} - e^{(c\theta)}}{e^{(-c\theta)} + be^{(c\theta)}}, \tag{4}$$

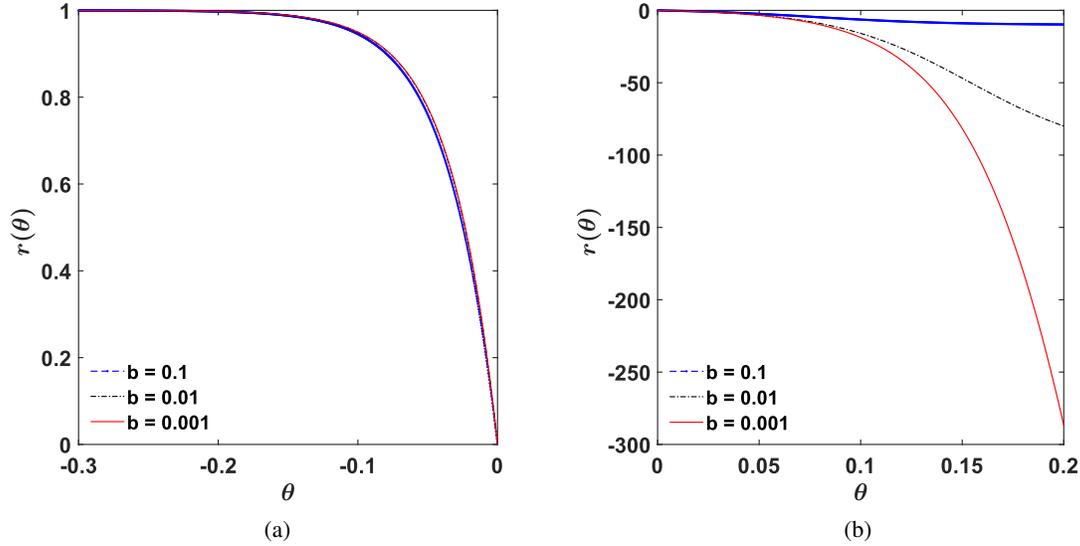

FIG. 2. Variation of the miscibility parameter $r$ with reduced temperature $\theta$: (a) Temperature below UCST i.e. $\theta < 0$ and (b) Temperature above UCST i.e. $\theta > 0$

where $c$ and $b$ are constants that are determined with the help of data available from experiments such as the density field and the surface tension. Fig. 2 depicts the variation of the miscibility parameter $r$ as a function of the reduced temperature $\theta$. The constants $c$ and $a$ holds relevance in the pre-UCST ($\bar{T} < \bar{T}_c \implies \theta < 0$) conditions, and are used to reproduce the experimentally observed variation of the thermophysical properties in the bulk regions of the fluid and the surface tension associated with the diffuse interface. For instance, the variation of density field for binary fluid pair FC72-1cSt silicone oil[50] can be reproduced by choosing $c = 10$ and $a = 0.6$, as previously demonstrated in our analysis of Rayleigh-Bénard-Marangoni convection[42]. In the pre-UCST limit, the value of the constant $b$ has negligible effect thanks to the formulation of the function $Z(\theta)$ as shown in fig. 2(a). However, the constant $b$ gains prominence in the post-UCST ($\bar{T} > \bar{T}_c \implies \theta > 0$) conditions and determines the rate of mass transfer across the interface (see fig. 2(b)). It is to be re-emphasized that even when the system temperature is higher than the UCST, the interface preserves its definition for a significant amount of time[31–34], i.e. as long as the concentration gradient is non-trivial. We have previously demonstrated the ability of the proposed function to reproduce the experimentally observed behavior in the context of FC72-1cSt (centiStokes) silicone oil[42]. From eq. 4, one may further deduce that while in the pre-UCST limit, for $\bar{T} << \bar{T}_c$ we get $r \to 1$, whereas in the post-UCST limit, i.e. for $\bar{T} >> \bar{T}_c$ we get $r \to \frac{-1}{b}$. The modified phase-field approach preserves the Korteweg stresses[6], which acts as long as the concentration gradient across the interface is large and mimics the effect of surface tension observed in immiscible fluids, in the post-UCST regime and therefore corroborates with the recent experimental findings of Truzzolillo et al.[33] and Carbonaro et al.[34].

The surface tension $\tilde{\sigma}$ associated with the diffuse interface is defined as the excess free energy per unit surface area[44]. Yue et al.[47] deployed the following relation for the calculation of surface tension for a one-dimensional planar interface:

$$\tilde{\sigma} = \int_{-\infty}^{\infty} [\tilde{f}_o(c) + \frac{\tilde{\Lambda}}{2}|\tilde{\nabla} c|^2] d\tilde{y}, \tag{5}$$

The spatio-temporal evolution of the flow field follows the second law of thermodynamics and the system of fluids is driven towards the state of minimization of the total free energy functional. The potential function responsible for the evolution of the flow field is termed as the chemical potential. The chemical potential $\tilde{\phi}$, defined as the variational derivative of the free energy functional, is defined as:

$$\tilde{\phi} = \frac{\delta \tilde{F}}{\delta c} = \tilde{f}_0'(c) - \tilde{\Lambda}\tilde{\nabla}^2 c = \frac{\tilde{\Lambda}}{\tilde{\epsilon}^2}\left(|r|^a c^3 - rc\right) - \tilde{\Lambda}\tilde{\nabla}^2 c, \tag{6}$$

The state of minimum total free energy functional is characterized by zero chemical potential ($\tilde{\phi} = 0$). This state is also called the state of thermodynamic equilibrium. Notably, for the system of fluids considered above UCST, the thermodynamic equilibrium can only be attained once the mixing process is complete which is marked by zero concentration gradient in the domain. However, for any system considered at a temperature below the UCST, one may obtain a well-defined steady state interface profile. The corresponding order parameter profile for a planar interface at an inhomogeneous state ($r > 0$) can be obtained from eq. 6 using $\tilde{\phi} = 0$ as follows[41]:

$$c_0 = -r^{\frac{(1-a)}{2}} \tanh\left(\frac{\tilde{y} - \tilde{y}_0}{\sqrt{2}(\tilde{\epsilon}/\sqrt{r})}\right), \tag{7}$$

where $\tilde{y}_0$ is the location of the interface. The effective interface thickness diverges as the miscibility parameter $r$ decreases

(or the system approaches the UCST) corroborating the findings of Buhn, Bopp, and Hampe [51] and Pousaneh, Edholm, and Maciolek [52]. The pre-factor $r^{\frac{(1-\alpha)}{2}}$ in eq. 7 ensures the initialization of correct profiles of the thermophysical properties in the domain by accounting for the degree of partial miscibility. As the degree of partial miscibility increases, the ratio of the individual thermophysical properties i.e. density ratio, viscosity ratio etc approaches unity. The consequence of the two effects - diverging interfacial thickness and miscibility induced redistribution of the concentration leads to the reduction of the surface tension which is indeed a naturally expected outcome of the increase in miscibility. Thus, for an interface in thermodynamic equilibrium, eqns. 5 and 7 can be used to obtain the surface tension as a function of the miscibility parameter $r$ as follows:

$$\tilde{\sigma} = \frac{2\sqrt{2}}{3} \frac{\tilde{\lambda}}{\tilde{\epsilon}} r^{\frac{(3-2a)}{2}} \qquad (8)$$

Eq. 8 can be rewritten as:

$$\tilde{\sigma} = \tilde{\sigma}_0 r^{\frac{(3-2a)}{2}}, \qquad (9)$$

where $\tilde{\sigma}_0$ is the surface tension at the limit of immiscibility i.e. $r \to 1$. Thus, the current formulation allows one to model both the immiscible fluids ($r = 1$) and the binary fluids with miscibility gap ($r < 1$). It is to be noted that the equations 7, 8 and 9 are obtained assuming thermodynamic equilibrium for $0 < r \leq 1$ and are therefore not applicable for the post-UCST regime ($r < 0$). The transition of the fluid pair, when heated to post-UCST condition is captured by the evolution of the bulk free energy density into a single well potential form and thus the surface tension exhibits a transient behavior, starting from the initially assumed surface tension at the immiscible limit [34]. While the nuances of the model are elucidated considering UCST fluids, it is imperative to mention that one may extend the formulation to model LCST fluids by just redefining the reduced temperature as $\theta = \frac{\tilde{T}_c - \tilde{T}}{\tilde{T}_c}$. The LCST fluids exhibit immiscibility or partial miscibility for system temperature $\tilde{T}$ higher than the critical temperature $\tilde{T}_c$. On an overall basis, we expect both UCST and the LCST fluid pairs to exhibit qualitatively similar behavior. Therefore, the rest of the analysis is performed considering UCST fluid pair.

## III. SYSTEM CONFIGURATION AND THE GOVERNING EQUATIONS

A system of binary fluids with UCST are considered in a superposed configuration in the presence of gravitational field as shown in fig. 3. The lighter fluid (fluid 2) with density $\tilde{\rho}_2$ lies over the heavier fluid (fluid 1) with density $\tilde{\rho}_1$ to avoid the manifestation of the Rayleigh-Taylor instability. Fig. 3(a) depicts the equilibrium profile of the interface as obtained from eq. 7. A tangent-hyperbolic velocity profile is imposed over the base density stratification to generate a shear flow at the interface. Subsequently, the interface profile and the imposed velocity profile is perturbed with a small-amplitude ($\tilde{h}_0$) periodic disturbance, with wavelength $\tilde{\lambda}$, to provoke the Kelvin-Helmholtz (KH) instability (see fig. 3(b)) following the work of Lee and Kim [53]. The wavelength $\tilde{\lambda}$ is considered to be equal to the lateral dimension of the domain $\tilde{L}$, where the later is chosen as the reference length scale in the current study. For simplicity [53], the height $\tilde{h}$ is chosen equal to the lateral dimension, i.e. $\tilde{h} = \tilde{L}$. Thus, the interface profile is initialized employing the following equation:

$$c(x, y, r) = -r^{\frac{(1-a)}{2}} \tanh\left(\frac{y - y_0 - h_0 sin(kx)}{\sqrt{2}(Cn/\sqrt{r})}\right), \qquad (10)$$

where $k(= \frac{2\pi}{\lambda})$ is the wavenumber of the imposed perturbation with $\lambda(= \tilde{\lambda}/\tilde{L})$ being the dimensionless wavelength and $Cn(= \tilde{\epsilon}/\tilde{L})$ is the Cahn number. Similarly, the imposed velocity profile is given as:

$$u(x, y) = \tilde{u}_{amp} \tanh\left(\frac{y - y_0 - h_0 sin(kx)}{\sqrt{2}(\tilde{\delta}_u/\tilde{L})}\right), \qquad (11)$$

where $\tilde{\delta}_u$ is the thickness of the velocity profile and $\tilde{u}_{amp}$ is the magnitude of the imposed velocity profile. It is to be noted that the initial velocity profile is unidirectional (in the lateral direction) in nature. The thicknesses of the velocity and the interface profiles are considered as independent parameters. One may define dimensionless velocity profile thickness as $\delta_u = \frac{\tilde{\delta}_u}{\tilde{L}}$. Figure 4 presents the variation of the order parameter for three distinct values of the miscibility parameter $r$ and velocity field for two distinct $\delta_u$ at the centerline ($x = 0.5$) along the height ($y$) of the domain. Further, to focus solely on the effect of miscibility on the flow dynamics, the initial velocity profile is considered independent of the miscibility parameter $r$. The imposition of the perturbation on the base equilibrium profile of the interface drives the system out of equilibrium ($\dot{\phi} \neq 0$). Further, the imposition of the shearing velocity profile, along with the tendency of the system to restore the thermodynamic equilibrium conditions, engenders a hydrodynamic flow. The subsequent spatio-temporal evolution of the order parameter field $c$ is governed by the advected form of the Cahn-Hilliard (CH) equation [46,54]. In current study, we employ the magnitude of the maximum initial velocity difference as the reference scale for the velocity i.e. $\tilde{u}_{ref} = |\tilde{U}_1 - \tilde{U}_2| = 2\tilde{u}_{amp}$ where $\tilde{U}_1, \tilde{U}_2$ are the maximum initial velocities in fluid 1 and fluid 2 respectively. The dimensionless CH equation as a function of the miscibility parameter $r$ can be expressed as:

$$\frac{\partial c}{\partial t} + \boldsymbol{\nabla} \cdot (\boldsymbol{u}c) = \frac{M}{Cn} \nabla^2 \left( |r|^a c^3 - rc - Cn^2 \nabla^2 c \right), \qquad (12)$$

where $t$ denotes time, and $M \left(= \frac{3}{2\sqrt{2}} \left(\frac{\tilde{\sigma}_0}{\tilde{u}_{ref} \tilde{L}^2}\right) \tilde{\gamma}\right)$ is the dimensionless measure of the interface mobility $\tilde{\gamma}$ and $\boldsymbol{u}$ represents the volume-averaged velocity field [43]. The mobility parameter should be chosen large enough to avoid the shear thinning of the interface due to the advective flow and simultaneously small enough to avoid over-damping of the flow[46,54]. The Navier-Stokes (NS) equations coupled with the continuity equation are employed to solve for the velocity ($\boldsymbol{u}$) and pressure ($P$) fields. The surface tension force, modelled along the lines of Korteweg stresses in the phase-field method, is accounted for in the NS equations following the approach of Jacqmin[46] and Yue *et al.*[47]. The continuity equation and the NS equation (reformulated as a function of the miscibility parameter $r$) are given as:

$$\boldsymbol{\nabla} \cdot \boldsymbol{u} = 0, \tag{13}$$

$$\frac{\partial (\rho_c \boldsymbol{u})}{\partial t} + \boldsymbol{\nabla} \cdot (\rho_c \boldsymbol{u} \otimes \boldsymbol{u}) = -\boldsymbol{\nabla} P + \frac{1}{Re} \boldsymbol{\nabla} \cdot \left( \mu_c \left( \boldsymbol{\nabla} \boldsymbol{u} + (\boldsymbol{\nabla} \boldsymbol{u})^T \right) \right)$$
$$+ \frac{1}{Fr^2} \rho_c g + \frac{3}{2\sqrt{2}} \frac{1}{WeCn} \left( |r|^a c^3 - rc - Cn^2 \nabla^2 c \right) \boldsymbol{\nabla} c$$
$$- \frac{1}{2} \frac{M}{Cn} \left( \boldsymbol{u} \left( 1 - \frac{\tilde{\rho}_2}{\tilde{\rho}_1} \right) \right) \nabla^2 \left( |r|^a c^3 - rc - Cn^2 \nabla^2 c \right) \quad (14)$$

where $Re (= \frac{\tilde{\rho}_1 \tilde{u}_{ref} \tilde{L}}{\tilde{\mu}_1})$ is the Reynolds number, $Fr(= \frac{\tilde{u}_{ref}}{\sqrt{\tilde{g}\tilde{L}}})$ is the Froude number and $We(= \frac{\tilde{\rho}_1 \tilde{u}_{ref}^2 \tilde{L}}{\tilde{\sigma}_0})$ is the Weber number. Equation 14 deserves further comment to elucidate the role played by the last two terms. The term containing $We$ and $Cn$ is responsible for the implementation of the surface tension force and is obtained by calculating $\tilde{\phi}\tilde{\boldsymbol{\nabla}}c$. The Laplacian term within the chemical potential helps to preserve the definition of the interface as long as the concentration gradient is non-zero. The last term in eq. 14 is introduced to ensure the thermodynamic consistency[48,55,56] of the model. This is primarily based on the work of Abels, Depner, and Garcke[55], wherein, the last convection term is introduced to extend the applicability of the model to high density ratio flows. Huang, Lin, and Ardekani[57] and Eikelder *et al.*[58] recently demonstrated the conservative nature of the overall formulation (eqns. 12 - 14). Further, $\rho_c$ and $\mu_c$ are order parameter based dimensionless density and dynamic viscosity, which can be written as:

$$\rho_c = \frac{\tilde{\rho}}{\tilde{\rho}_1} = \left(\frac{1+c}{2}\right) + \frac{\tilde{\rho}_2}{\tilde{\rho}_1}\left(\frac{1-c}{2}\right), \tag{15}$$

$$\mu_c = \frac{\tilde{\mu}}{\tilde{\mu}_1} = \left(\frac{1+c}{2}\right) + \frac{\tilde{\mu}_2}{\tilde{\mu}_1}\left(\frac{1-c}{2}\right), \tag{16}$$

with $\mu_1$ and $\mu_2$ being the dynamic viscosity of the heavier and the lighter fluid respectively.

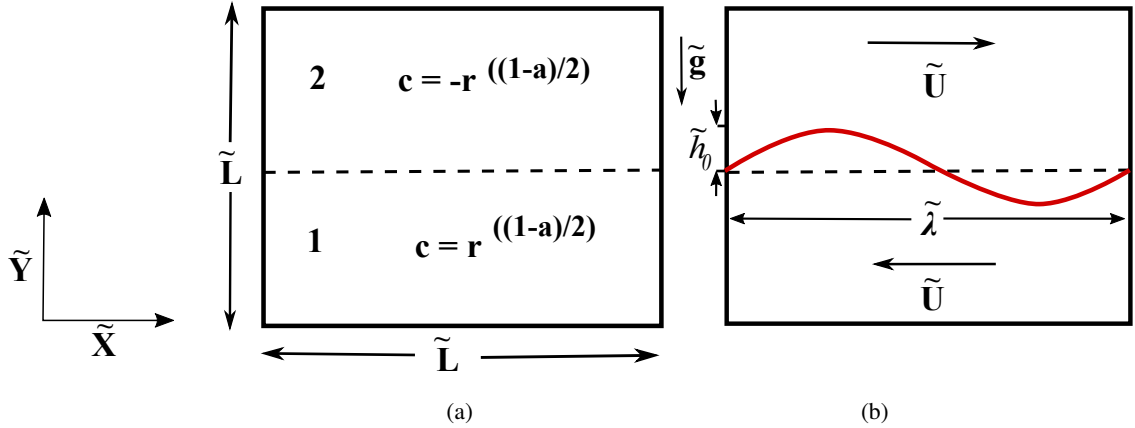

(a)                                    (b)

FIG. 3. Schematic representation of the fluids configuration: (a) Unperturbed fluid setup with lighter fluid 2 lying over the heavier fluid 1 and (b) Imposed shear flow over the perturbed interface

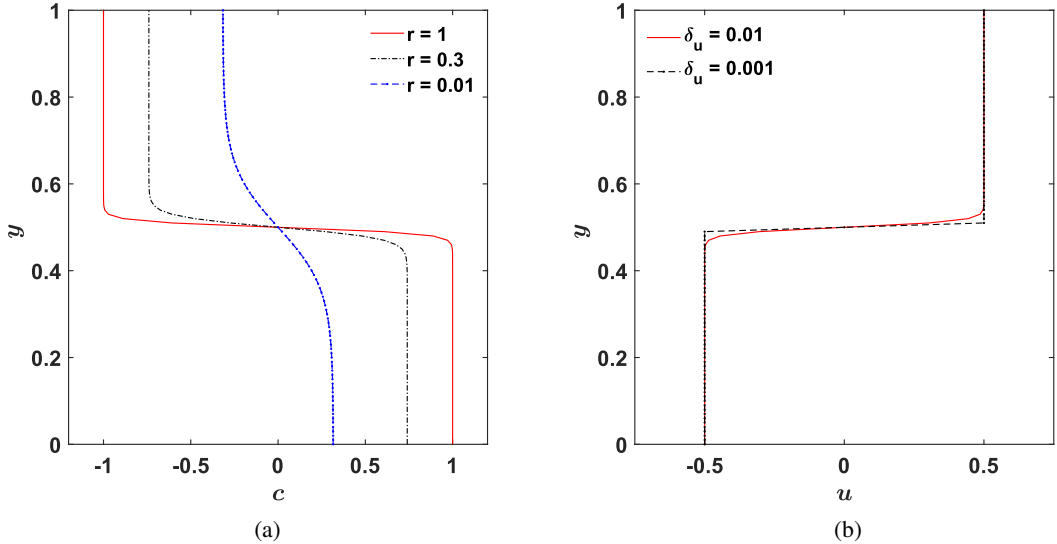

FIG. 4. (a) Initial order parameter and (b) velocity profile

## IV. NUMERICAL FORMULATION

Within the purview of binary fluids exhibiting temperature sensitive miscibility gap, the dynamics of the isothermal KH instability can be explored in two different configurations on the basis of thermodynamic equilibrium as shown in fig. 5. In the first case, hereafter referred as case (*I*), the two fluids are allowed to attain thermodynamic equilibrium corresponding to the system temperature. Subsequently, the shearing velocity profile along with perturbation to the interface (as determined from eqns. 10-11) is imposed. This case is limited to a system temperature below UCST (see fig. 5(a)) as there will be no equilibrium interface profile for system of fluids considered above UCST . The second configuration on the other hand, hereafter referred as case (*II*), deals with thermodynamic non-equilibrium cases. The temperature of the system of fluids is raised instantaneously to a temperature higher than the UCST (see fig. 5(b)), thereby additionally allowing mass transport across the interface. We assume the two fluids to be incompressible and Newtonian in nature. The geometrical dimensions of the domain is considered as [$L, L$]. Further, the thermophysical properties of the fluids pertinent to the study namely density and viscosity, are assumed to exhibit indirect dependence on temperature through the miscibility parameter $r$. Thus, depending upon the temperature of the system, the thermophysical properties are governed by eqns. 15 and 16 with the order parameter being governed by eq. 10. The finite volume formulation is employed for discretization of the governing equations and the resulting algebraic equations are then solved over uniform cartesian mesh. We have modified the open-source solver phaseFieldFoam[59] in order to solve the set of equations. The details on the numerical methodology are presented in the ensuing section.

## A. Numerical methodology

In our previous study[41], we extended the phaseFieldFoam[43,59] solver to resolve the miscibility parameter dependent governing equations in the context of isothermal Rayleigh-Taylor instability. The phaseFieldFoam is developed on the OpenFoam framework where the equations are discretized using second-order finite-volume scheme. The coupled Cahn-Hilliard-Navier-Stokes (CH-NS) equations are solved in a segregated manner. The PIMPLE algorithm, which is a combination of PISO (Pressure implicit with splitting of operator)[60] and SIMPLE (Semi-implicit method for pressure linked equations)[61], is employed to couple the velocity and pressure fields.

In the beginning of each iteration of the PIMPLE algorithm, the CH equation (eq. 12) is solved to update the order parameter field $c$. Subsequently, the Korteweg stress term of the momentum equation (eq. 14) is calculated. Finally the mass and momentum conservation equations (eqns. 13-14) are solved to update the velocity and pressure fields. The updated velocity field is then utilized to repeat the iterative cycle. It is to be noted that the CH equation involves discretization of a fourth-order term and is therefore very challenging. An inaccurate discretization would result in mass accumulation in high-velocity regions. Thus, to avoid the development of associated spurious currents, the CH equation is split into two second-order equations during the first step of PIMPLE algorithm. The chemical potential $\tilde{\phi}$ is first solved and is subsequently substituted in the diffusion term of the CH equation (eq. 12) to update the order parameter field $c$.

Yue *et al.*[47] have summarized the challenges associated with the phase-field based formulations. One such important challenge is the numerical oscillations. Given the dependence of the Korteweg stresses on the concentration gradient, it is important to preserve the local sharpness of the interface and minimize numerical diffusion. Thus, the advective terms in the CH equation (eq. 12) and the momentum equation (eq. 14) are discretized using the Gauss Gamma[62] and Gauss limited linear V[63] schemes, respectively, both of which are high-resolution, nonlinear NVD (Normalized Variable Diagram) and TVD (Total Variation Diminishing) schemes[64]. On the other hand, the diffusive terms in these equations are discretized using the Gauss linear method, and the temporal terms are handled using the first-order implicit Euler time-stepping scheme[65]. Subsequently, preconditioned bi-conjugate gradient (BiCG) method of Van-Der-Vorst[66] is used to iteratively solve the coupled equations.

## B. Initial and boundary conditions

Initially (at $t = 0$), a velocity profile given by eq. 11 is imposed over the base order parameter profile (eq. 10). The wavenumber of the perturbation $k (= 2\pi)$ is kept constant throughout the current study. For the simulations pertaining to case (*I*), the system is initialized at distinct values of the miscibility parameter $r (= 0.01, 0.3$ and $1)$, i.e., the fluid pair is allowed to attain a state of thermodynamic equilibrium. The miscibility parameter $r$ is held constant during the course of the simulation. Considering three distinct values of $r$ allows us to explore the effect of relative proximity of the system temperature to the UCST. Whereas for the simulations pertaining to case ($II$), the value of the miscibility parameter $r$ is changed instantaneously to $r < 0$ thereby signifying a system temperature above UCST which results in thermodynamic non-equilibrium. Consequently, the transformation of the bulk free energy density from a double well potential to a single well potential drives the mass transfer across the interface. Therefore, case ($II$) investigates the interplay between developing KH instability and simultaneous mass transport. The governing equations (eqns. 12-14) are subjected to periodic boundary conditions in the lateral direction and zero-gradient boundary conditions in the vertical direction[38] except for the vertical component of the velocity. The top and bottom walls of the domain are assumed to be impenetrable and therefore $v = 0$ at $y = 0, 1$ where $v$ is the vertical component of the velocity field $\boldsymbol{u}$.

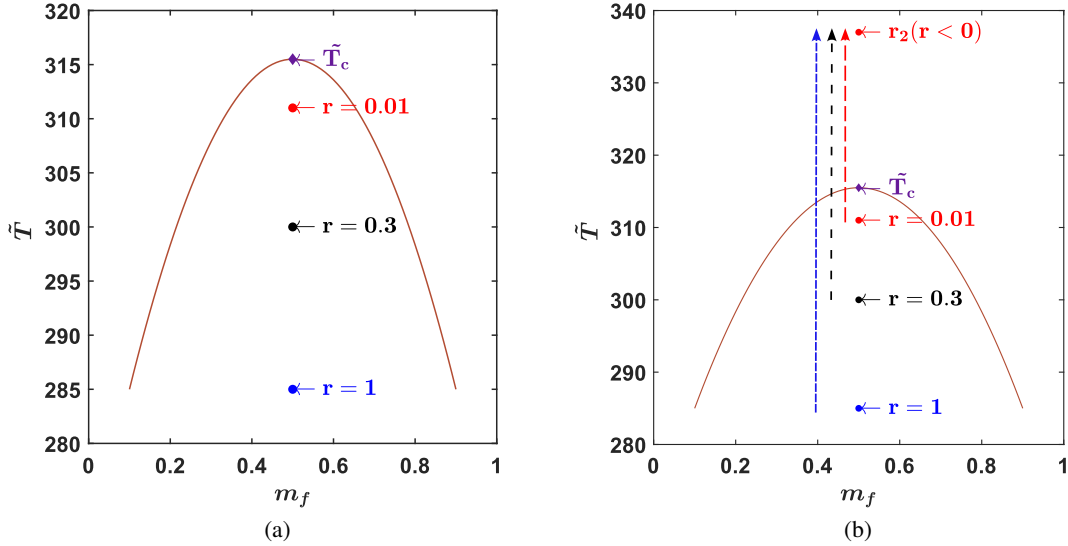

FIG. 5. Schematic representation of distinct configurations of analysis: (a) Thermodynamic equilibrium and (b) Thermodynamic non-equilibrium. The temperature values are chosen only to prepare the schematic and correspond to binary fluid pair of FC72-1cSt silicone oil. It is to be noted that different binary fluid pairs have different UCST.

### C. Validation

The current numerical solver has been extensively validated in the context of immiscible fluid pairs[43,59]. In our previous study[41], we performed a thorough comparison of linear and non-linear results for RT instability for binary fluids exhibiting UCST. We observed a relative error of less than 2% between linear and non-linear analysis, thereby establishing the accuracy and the validity of our model. Here, we present a comparison of the results from our numerical solver with the benchmarks results of Lee and Kim[53] in the context of KH instability in immiscible fluids, i.e. $r = 1$. The thermophysical properties of the fluids are determined through the pertinent dimensionless numbers namely Reynolds number ($Re$), Weber number ($We$) and the density ratio ($\rho_r = \tilde{\rho}_2/\tilde{\rho}_1$). The viscosity ratio ($\mu_r = \tilde{\mu}_2/\tilde{\mu}_1$) is considered unity throughout the current analysis. The geometrical parameter associated with the phase-field method namely the interface thickness is governed through Cahn number ($Cn$) which is defined considering the interfacial thickness at the limit of immiscibility ($r = 1$). The $Cn (= 0.01)$ is kept constant throughout the present analysis. The dimensional mobility ($\tilde{\gamma}$) follows the relation $\tilde{\gamma} = 0.01\tilde{\epsilon}^2$ as suggested by Jamshidi $et~al.$ [59]. Further, for the validation tests, the Froude number $Fr$ is also considered unity[53]. To begin with, a thorough grid convergence test is conducted to obtain the optimum grid size that is utilized to generate the forthcoming results. The dimensionless numbers pertinent to the fluids and the flow configuration are $Re = 5000$, $We = 10^5$ and $\rho_r = 0.99$. Four different grid sizes namely $250 \times 250$, $500 \times 500$, $750 \times 750$ and $1250 \times 1250$ are considered. A qualitative comparison of the interface topology for different grid sizes is presented in fig. 6. Only three among the four considered grid sizes are presented for brevity. As clearly evident, the coarser grid ($250 \times 250$) fails to capture the tortuous interface topology while the finer grids ($750 \times 750$ and $1250 \times 1250$) resolves the same. To quantify the effects, we compute the temporal variation of the billow height and the kinetic energy. The billow height $\tilde{H}_B$ is defined as the vertical distance of separation between the maximum and the minimum height taken by the interface during the development of the KH instability. The dimensionless billow height can therefore be defined as $H_B = \tilde{H}_B/\tilde{L}$. The kinetic energy $\widetilde{E_k}$ is calculated by employing the following relation:

$$\widetilde{E_k} = \tilde{\rho}_1 \tilde{u}_{ref}^2 \left[ \sum \frac{1}{2} \rho_c u^2 \right], \tag{17}$$

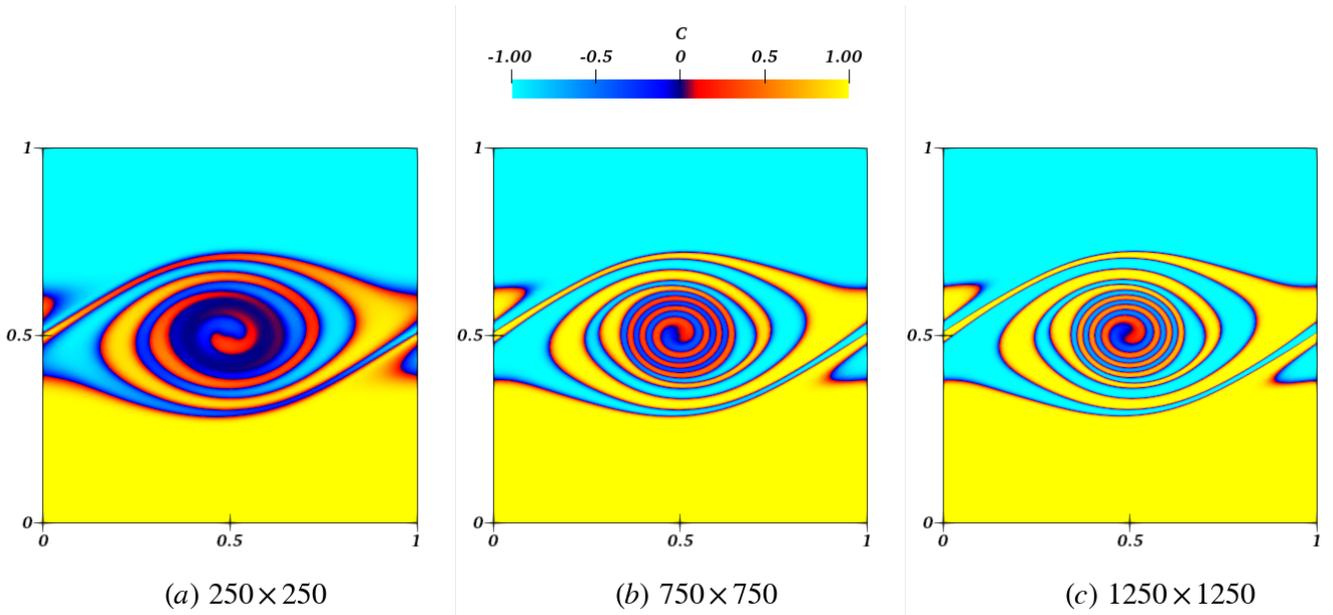

FIG. 6. Qualitative grid independence test - Comparison of interface topology at $t = 1.4$ for (a) $250 \times 250$, (b) $750 \times 750$ and (c) $1250 \times 1250$. The rest of the pertinent dimensionless parameters are $\rho_r = 0.99$, $Re = 5000$, $We = 10^5$ and $Fr = 1$.

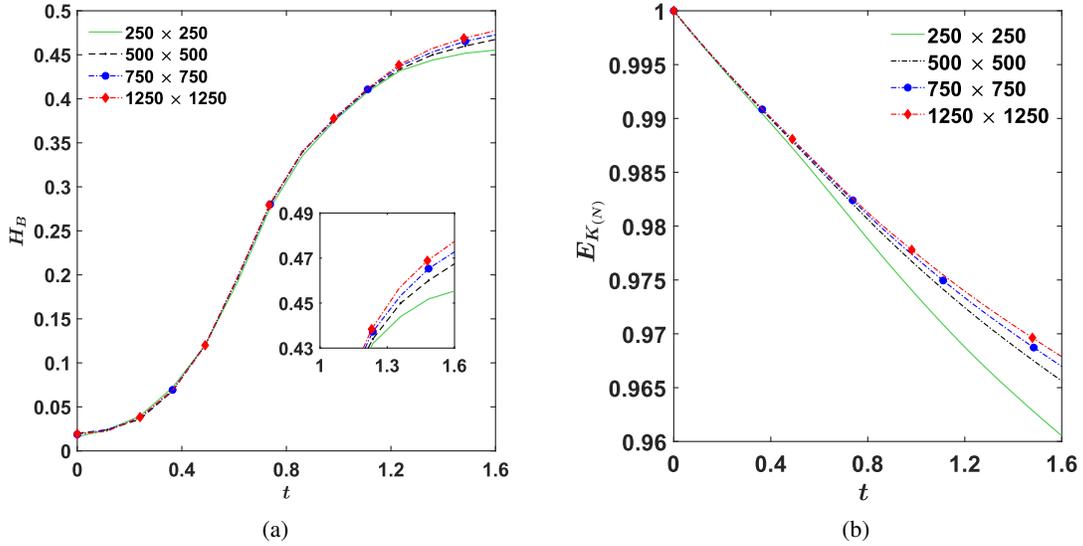

FIG. 7. Quantitative grid independence test - Comparison of (a) Billow Height and (b) Normalized kinetic energy. The rest of the pertinent dimensionless parameters are $\rho_r = 0.99$, $Re = 5000$, $We = 10^5$ and $Fr = 1$.

Subsequently, the normalized kinetic energy $E_{K_{(N)}} (= \widetilde{E}_{K_{(t)}} / \widetilde{E}_{K_{(0)}})$ is calculated. Fig. 7 depicts the temporal variation of the billow height and the normalized kinetic energy. The billow height is found to be relatively insensitive to the grid size as no difference was observed till $t = 1.2$ for all the grids considered. The kinetic energy, on the other hand, is highly sensitive to the grids and is therefore a better-suited physical quantity to be compared while performing grid-convergence test. Thus, on the basis of balance between the desired accuracy and the computational costs, a $750 \times 750$ mesh is chosen as the optimum grid size. Finally we present the comparison of our results with the benchmark results of Lee and Kim[53] in fig. 8. The excellent agreement between the two solutions establishes the reproducibility of the results presented in the ensuing section.

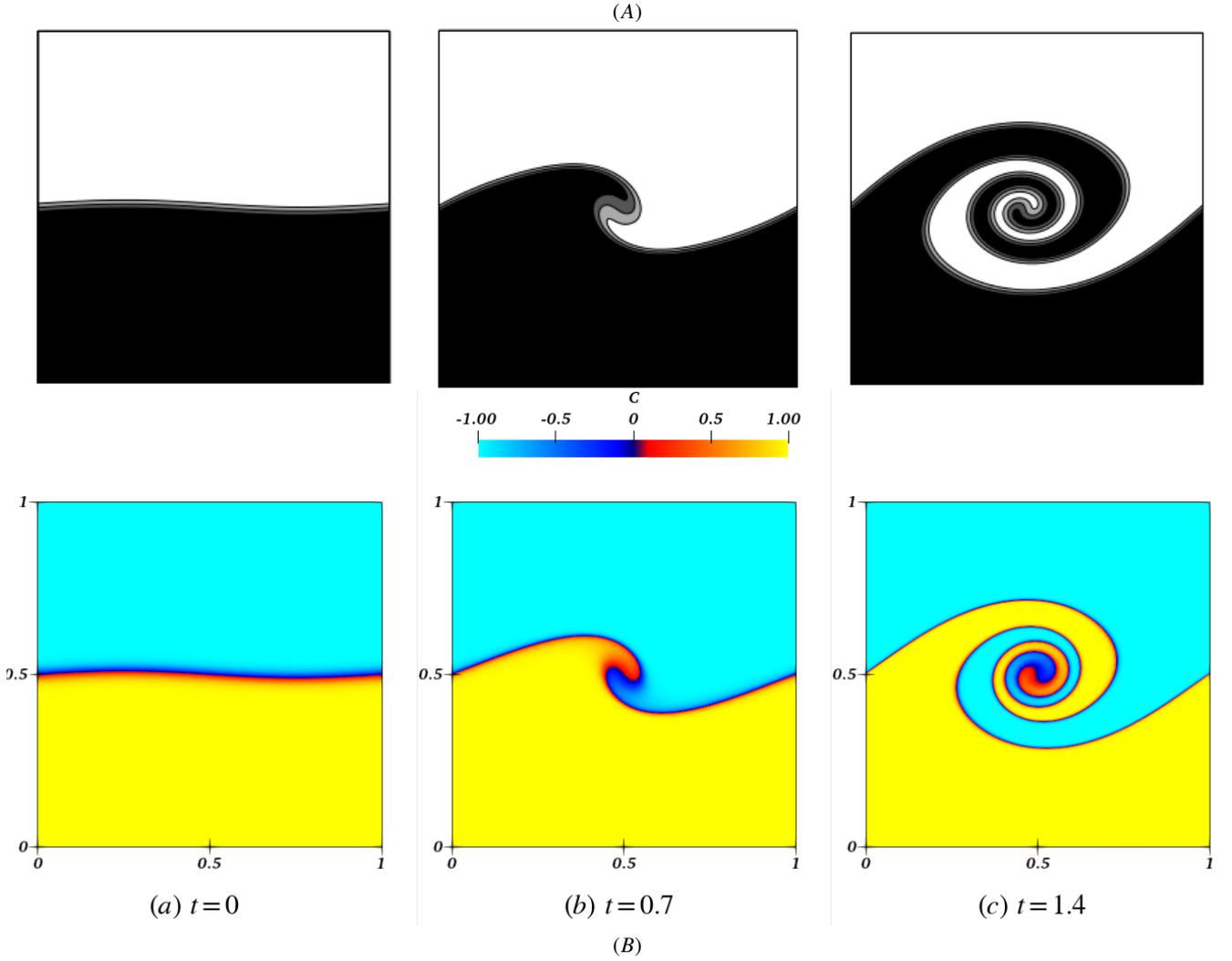

FIG. 8. Validation of the current solver against the benchmarks results[53] for the temporal evolution of the interface: (*A*) Lee and Kim [53] and (*B*) current simulations. The rest of the pertinent dimensionless parameters are $\rho_r = 0.99$, $Re = 5000$, $We = 10^5$ and $Fr = 1$.

## V. RESULTS AND DISCUSSION

The results of the numerical simulations pertaining to KH instability is analyzed by performing a thorough energy budget analysis. The calculation of normalized kinetic energy has been demonstrated in the previous section (see eqn. 17). The surface energy and the potential energy of the system are computed using the following relations:

$$\widetilde{S.E.} = \tilde{\rho}_1 \tilde{u}_{ref}^2 \left( \frac{3}{2\sqrt{2}} \frac{Cn}{We} \left[ \sum |\nabla c|^2 \right] \right), \tag{18}$$

$$\widetilde{P.E.} = \tilde{\rho}_1 \tilde{g} \tilde{L} \left[ \sum \rho_c g y \right]. \tag{19}$$

One may further compute the dimensionless normalized surface energy $E_{S_{(N)}}$ and the potential energy $E_{P_{(N)}}$ as follows:

$$E_{S_{(N)}} = \frac{\widetilde{E}_{S_{(t)}}}{\widetilde{E}_{S_{(0)}}}, \tag{20}$$

$$E_{P_{(N)}} = \frac{\widetilde{E}_{P_{(t)}}}{\widetilde{E}_{P_{(0)}}}, \tag{21}$$

As previously stated, the analysis is further divided into two subsections based on the thermodynamic state of the system. Since the novelty of the work lies in incorporation of a tunable miscibility framework through the parameter $r$, the density ratio ($\rho_r = \tilde{\rho}_2/\tilde{\rho}_1$) is held constant throughout the analysis as the role of density ratio in the KH instability is well explored in the literature. Typically, the density ratio for binary fluids with miscibility gap ranges from 0.45 to 0.95. Therefore, we have chosen $\rho_r = 0.8$ unless stated otherwise. In this study, we first analyze the effect of Weber number $We$, the effect of velocity profile thickness $\delta_u$ and the effect of the stratification by changing Froude number $Fr$. Subsequently, in $II$ case, the fluid pair is instantaneously heated to provoke interfacial mass transport. A detailed analysis of the influence of the initial and final temperature of the system, controlled through the miscibility parameter $r$ is next presented.

## A. Initialization with thermodynamic equilibrium

We begin our analysis under the equilibrium configuration by considering three different values of the miscibility parameter $r(= 0.01, 0.3, 1)$. The three distinct values represent the successively decreasing proximity of the system temperature with the UCST of the binary fluids. As noted previously, $r = 1$ refers to the theoretical limit of immiscibility in the context of binary fluids. Fig. 9 demonstrates the initial condition (recall sec. III). Evidently, the interfacial thickness diverges as the value of the miscibility parameter is reduced, i.e. as the system approaches the UCST corroborating the findings of Buhn, Bopp, and Hampe[51] and Pousaneh, Edholm, and Maciolek[52]. The homogenization of the order parameter field (or the density field) due to the miscibility can be quantified by computing the variance[30] employing the following expression:

$$\chi = \left( \frac{\partial \rho_c}{\partial x_i} \frac{\partial \rho_c}{\partial x_i} \right) \qquad (22)$$

### 1. Effect of Weber number

The spatio-temporal evolution of the interface subjected to shearing motion is strongly dependent on the surface tension of the fluid pair. Thus, in order to quantify the effect of the surface tension, three different values of the Weber number $We(= 100, 1000$ and $10000)$ are considered. It is to be re-emphasized that the Weber number is defined using the surface tension values at the limit of immiscibility. The subsequent evolution of the surface tension owing to the temperature induced miscibility is accounted through the miscibility parameter $r$. While the surface tension is indirectly dependent on temperature due to miscibility, we assume no direct temperature dependence thereby neglecting the Marangoni flow. The rest of the parameters pertinent to numerical simulations are $Re = 1000$, $\delta_u = 0.01$ and $Fr = 1$. Fig. 10, 11 and 12 present the interface topology at $t = 8$ for $We = 100, 1000$ and $10000$ respectively for the three values of the miscibility parameter $r(= 0.01, 0.3, 1)$. As the temperature of the system approaches the UCST i.e. the value of $r$ decreases, the enhanced miscibility leads to the reduction of surface tension, effective density ratio and increment of the interfacial thickness. Consequently, the absolute values of the kinetic energy, potential energy and the surface energy is a function of $r$. One needs to simultaneously follow the energy budget and variance analysis alongside the topological snapshots (see fig. 10-12) to understand the flow behavior. The temporal evolution of the normalized kinetic energy, potential energy and the surface energy are shown in fig. 13(a,b and c) respectively. Fig. 13(d) plots the variance of the density field.

To begin with, let us consider the case of $We = 100$. Our analysis reveals that for early time stages, i.e. till $t = 2$, all the three case exhibit qualitatively similar behavior. The interface distorts due to the imposed shearing mechanism. However, owing to the relatively higher surface tension at the limit of immiscibility, the interface in $r = 1$ (see fig. 10(c)) case retracts towards the initial profile. The surface energy analysis corroborates this observation as shown in fig. 13(c). As the value of the parameter $r$ is decreased, the enhanced mixing prior to the shearing results in weaker stratification. Thus, the fluid pair is more susceptible to the manifestation of the KH instability. For $r = 0.3$, a finger-like protrusion of one fluid into the other is observed. The protrusion is eventually deformed due to the base flow in both layers of the fluids leading to the generation of an almost horizontal section of the fluid as shown in fig. 10(b). On the other hand, the case $r = 0.01$ leads to the formation of a stationary vortex which causes the billow formation. Thus, the rate of increase of surface energy is highest for $r = 0.01$ (see fig. 13(c)). The analysis of normalized kinetic and potential energy (see fig. 13(a,b)) reveals interesting insights of the process. At early time stages, the rate of fall of the kinetic energy is smaller for smaller value of the parameter $r$. However, at $t = 3.5$, the $E_{K_N}$ for $r = 0.3$ becomes lower than that of $r = 1$, whereas the $r = 0.01$ case continues following the trend of slower fall of the kinetic energy. Similarly, the normalized potential energy follows a non-monotonic trend. The normalized potential energy for $r = 0.3$ increases higher as compared to $r = 1$ and $r = 0.3$. Such behavior is expected as the initial kinetic energy is the driving potential for the KH instability. The surface energy and the potential energy rise at the expense of the kinetic energy. Thus, one can argue for the existence of a threshold degree of partial miscibility till which the rise in potential energy trips the case of completely immiscible configuration. Subsequently, as the degree of partial miscibility rises, the rate of increase in the potential energy and the absolute increase of the potential energy falls. The temporal evolution of the potential energy may further be explained by comparing the variance of the density field as shown in fig. 13(d). The initial variance of the density field decreases monotonically as the degree of partial miscibility is increased (or the miscibility parameter $r$ is decreased). It is to be re-emphasized that the initial configuration has a stable density stratification in order to avoid manifestation of Rayleigh-Taylor instability. Consequently, the potential energy which is a function of local fluid density and elevation is lowest for immiscible case ($r = 1$). As the system of fluids approach the UCST, the enhanced mixing leads each fluid layer having an effective r-averaged density. Thus, the mixing results in a comparative increase in the density of the upper layer of the fluid and a decrease in the density of the lower layer of the fluid leading to a rise in the initial potential energy of the system. Since, the case $r = 1$ leads to retraction of the interface, the normalized potential energy falls after the initial rise. This results in an oscillatory rise of the overall potential energy of the system unlike $r = 0.01$. Finally, the advection-induced thinning of the interface for $r = 0.3$ and $0.01$ increases

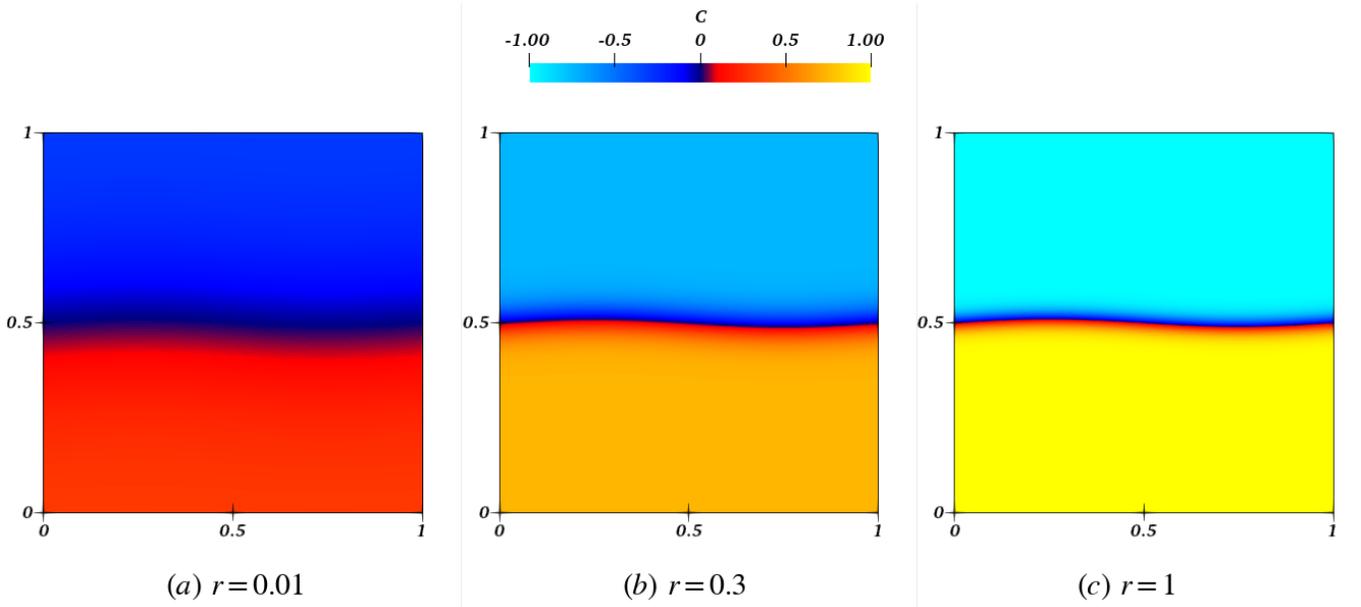

FIG. 9. Initial Condition: (a) $r = 0.01$, (b) $r = 0.3$ and (c) $r = 1$

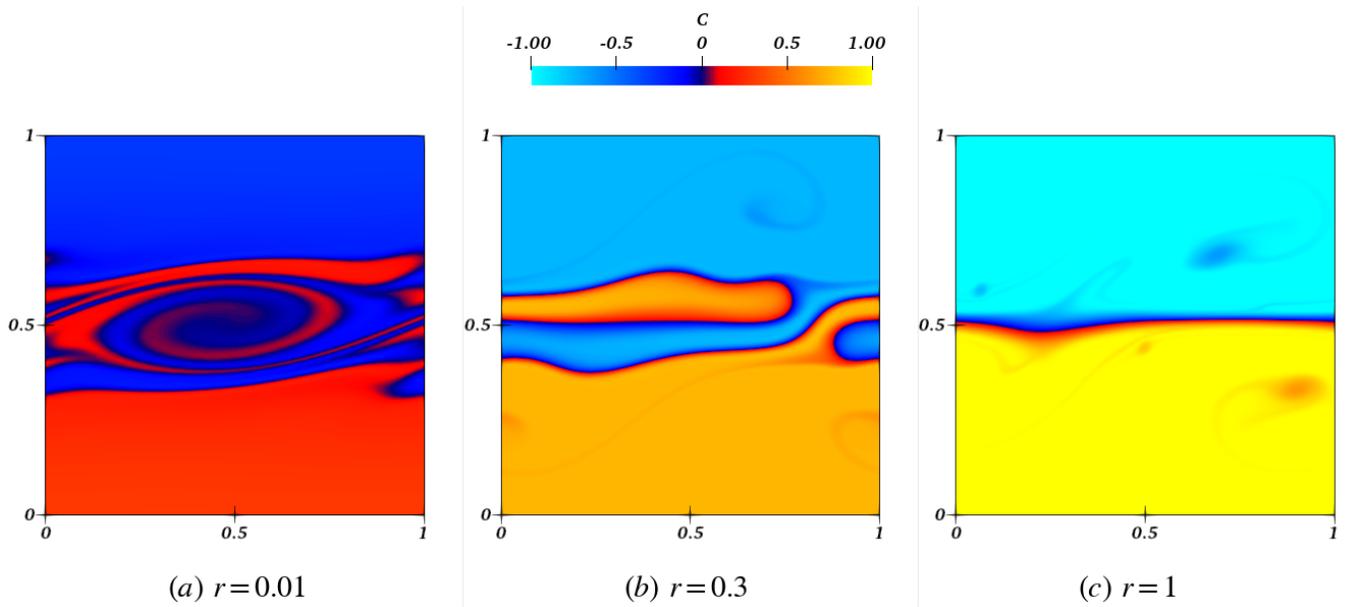

FIG. 10. Interface topology at $t = 8$ for $We = 100$ for three distinct values of the miscibility parameter $r$: (a) $r = 0.01$, (b) $r = 0.3$ and (c) $r = 1$. The rest of the parameters pertinent to numerical simulations are $\rho_r = 0.8$, $Re = 1000$, $\delta_u = 0.01$ and $Fr = 1$.

the variance in the domain.

On the other hand, if the initial surface tension is relatively lower i.e. $We = 1000$ or $10000$, all three values of the miscibility parameter leads to billow formation. Interestingly, while the vortex is stationary for $r = 0.01$, it is traveling for $r = 0.3$ and 1. Thus, the degree of partial miscibility also leads to demarcation of boundary between the KH instability and the Homlboe instability. For higher values of the Weber number, a monotonic trend is observed in the energy budget analysis. The rise in potential energy is now highest for $r = 1$. The billow formation is associated with a higher rise in the potential energy of the system. That being said, the initial density disparity between the two layers is another important parameter. Higher the initial density disparity, higher would be the rise of potential energy during billow formation. Therefore, we get a monotonic rise of the normalized potential energy as $r$ decreases. For $r = 0.01$, all three $We$ leads to billow formation as the mixing leads to significant reduction in density stratification which is the primary stabilizing agent. The reduced variance in this case however leads to very low density disparity and hence the potential energy rise is much lower as compared to $r = 0.3$ and 1. Therefore, the $We$ number in unison with the miscibility parameter $r$ plays a key role in not only governing the dynamics of the roll formation but also in determination of threshold system temperatures to allow the manifestation of KH instability.

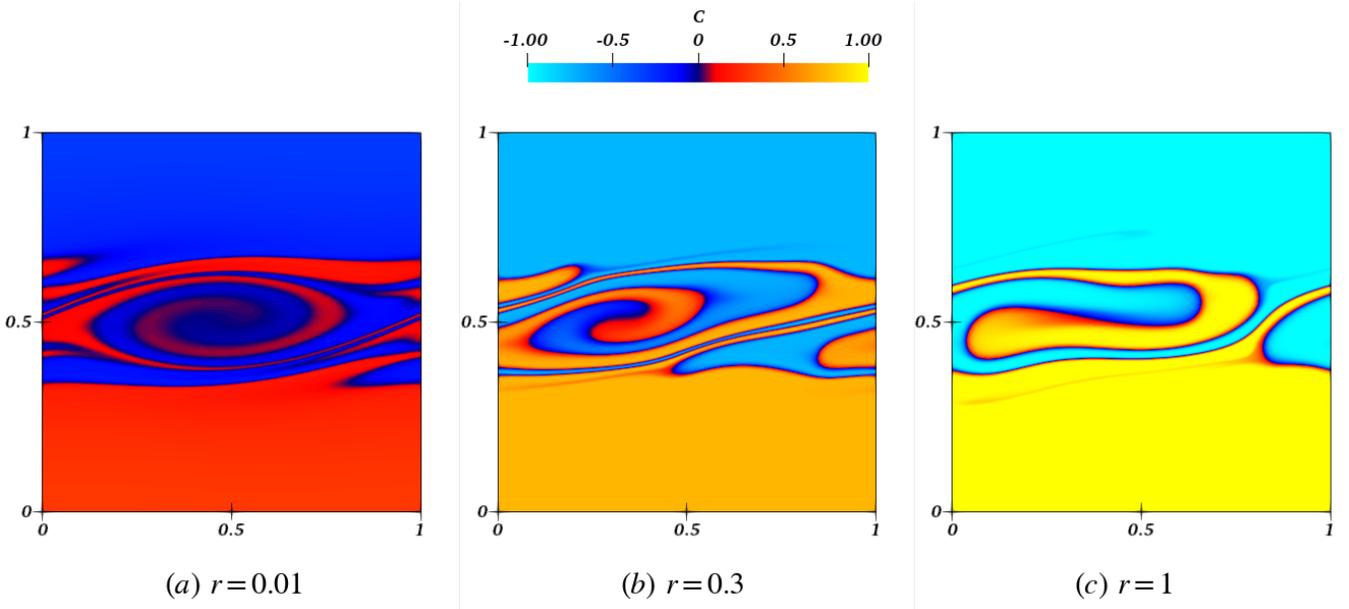

(a) $r = 0.01$     (b) $r = 0.3$     (c) $r = 1$

FIG. 11. Interface topology at $t = 8$ for $We = 1000$ for three distinct values of the miscibility parameter $r$: (a) $r = 0.01$, (b) $r = 0.3$ and (c) $r = 1$. The rest of the parameters pertinent to numerical simulations are $\rho_r = 0.8$, $Re = 1000$, $\delta_u = 0.01$ and $Fr = 1$.

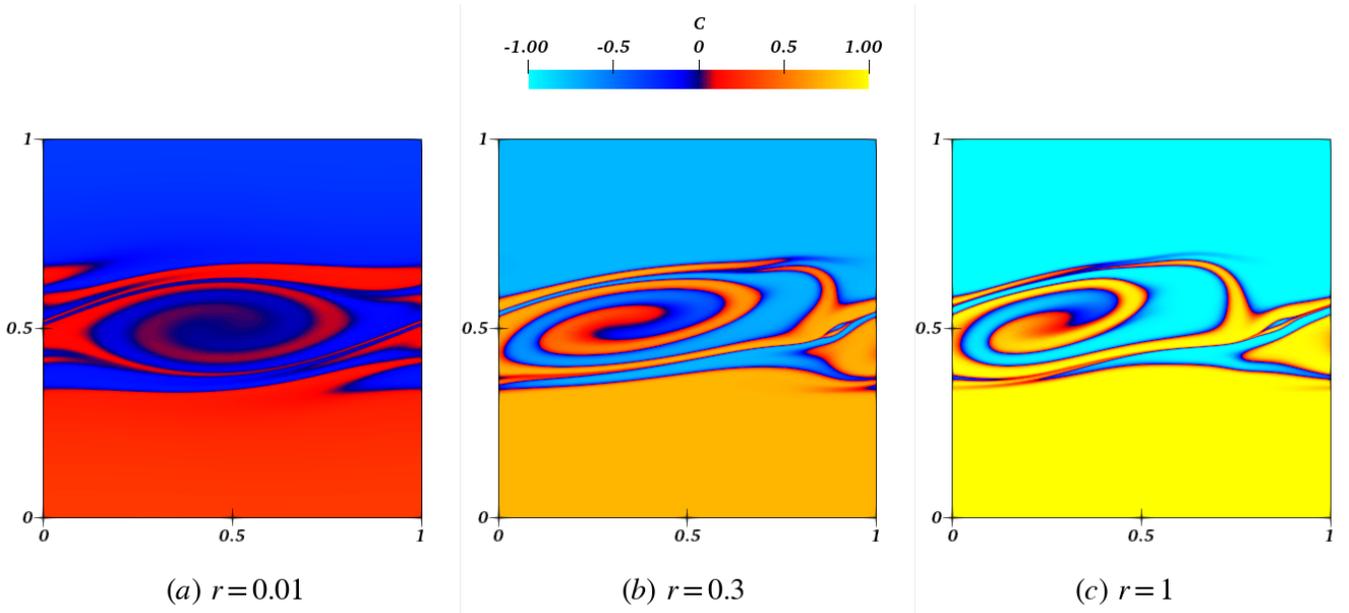

(a) $r = 0.01$     (b) $r = 0.3$     (c) $r = 1$

FIG. 12. Interface topology at $t = 8$ for $We = 10000$ for three distinct values of the miscibility parameter $r$: (a) $r = 0.01$, (b) $r = 0.3$ and (c) $r = 1$. The rest of the parameters pertinent to numerical simulations are $\rho_r = 0.8$, $Re = 1000$, $\delta_u = 0.01$ and $Fr = 1$.

## 2. Effect of velocity profile thickness

The relative length scales of the velocity and the density profiles have been found to influence the onset of KH instability[19]. Since the temperature induced miscibility leads to the divergence of the interfacial thickness, it is imperative to investigate the combined effect of the degree of miscibility and the velocity profile thickness ($\delta_u$). Having already investigated the case of $\delta_u = 0.01$ in the previous section, we performed nu-merical simulations for $\delta_u = 0.001$. The rest of the parameters pertinent to numerical simulations are $\rho_r = 0.8$, $Re = 10000$, $We = 1000$ and $Fr = 1$. Fig. 14 presents the interface topology for $\delta_u = 0.001$ at $t = 2$ and $t = 8$. The billow formation is observed for all values of the miscibility parameter $r$. While $r = 0.01$ lead to the formation of a stationary vortex, $r = 0.3$ and $1$ result in a traveling vortex. Thus, both thicknesses of the velocity profile produced qualitatively similar results. However, marginal quantitative differences are observed as summarized in fig. 15. While figs. 15(a,b and c) depict the normalized kinetic, potential and the surface energy, fig. 15(d) presents the variance of the density field as a function of time. The sharper velocity profile induces faster advection based sharpening of the interface at early time stages. Consequently, the variance in the case of $\delta_u = 0.001$ rises higher as compared to $\delta_u = 0.01$ at early time stages(see fig. 15(d)). Now since the Korteweg stresses in the phase-field method are implemented through the gradient of the order parameter profile, an increase in variance at early time stages results in correspondingly higher surface energy as also evident in fig. 15(c). Interestingly, only marginal changes are observed in the normalized potential energy. Further, the normalized kinetic energy profile follows a monotonic trend throughout the interval of the simulations with $\delta_u = 0.001$ having higher fall as compared to $\delta_u = 0.01$. The higher loss of kinetic energy is also attributed to the higher dissipation in case of sharper velocity profile. In conclusion, the sharpness of the velocity profile follows similar trend irrespective of the partial degree of miscibility between the two fluids.

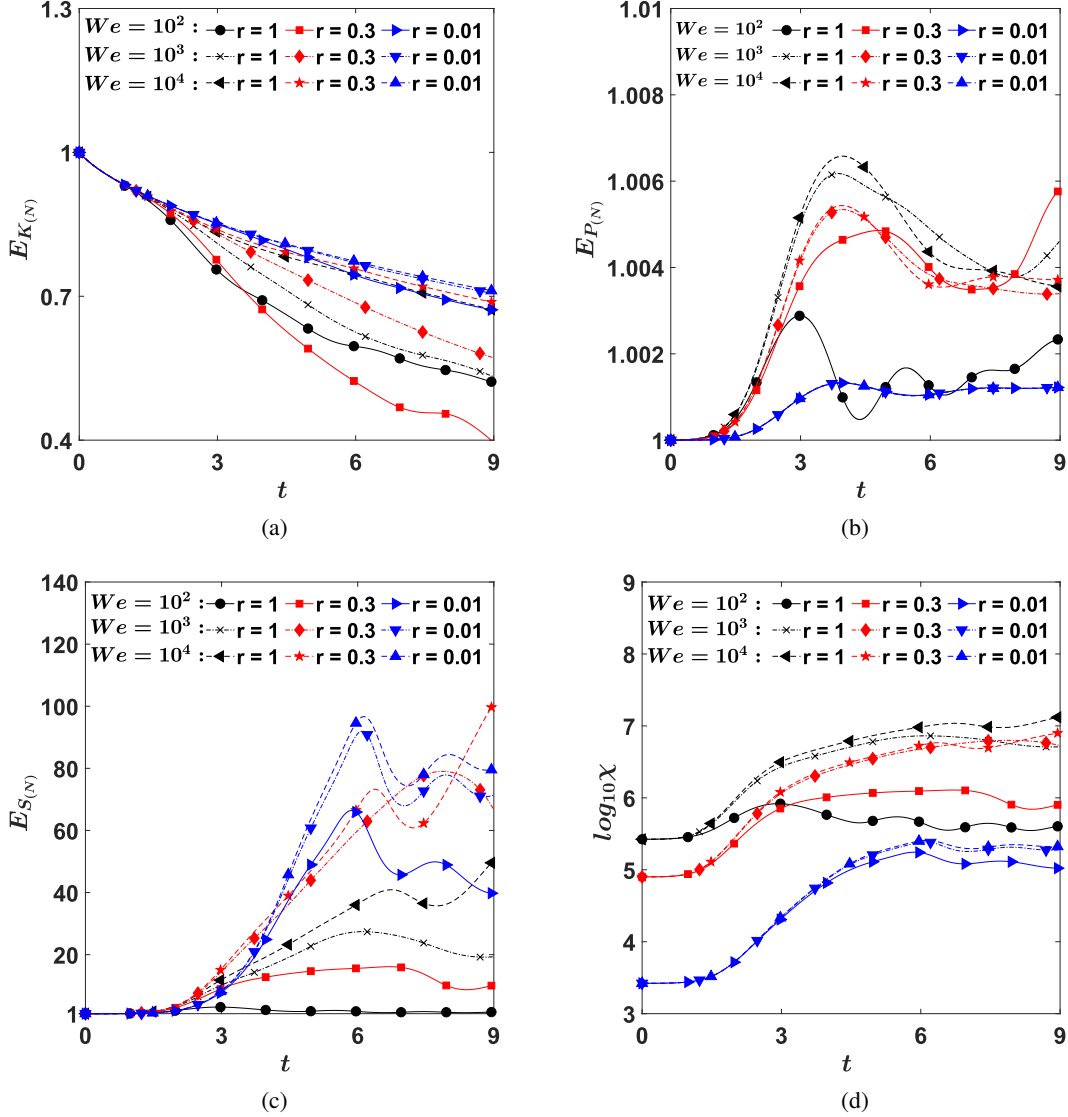

FIG. 13. Quantitative comparison of the effect of Weber number and degree of miscibility: (a) Normalized kinetic energy, (b) normalized potential energy, (c) normalized surface energy and (d) Variance of the density field. The rest of the parameters pertinent to numerical simulations are $\rho_r = 0.8$, $Re = 1000$, $\delta_u = 0.01$ and $Fr = 1$.

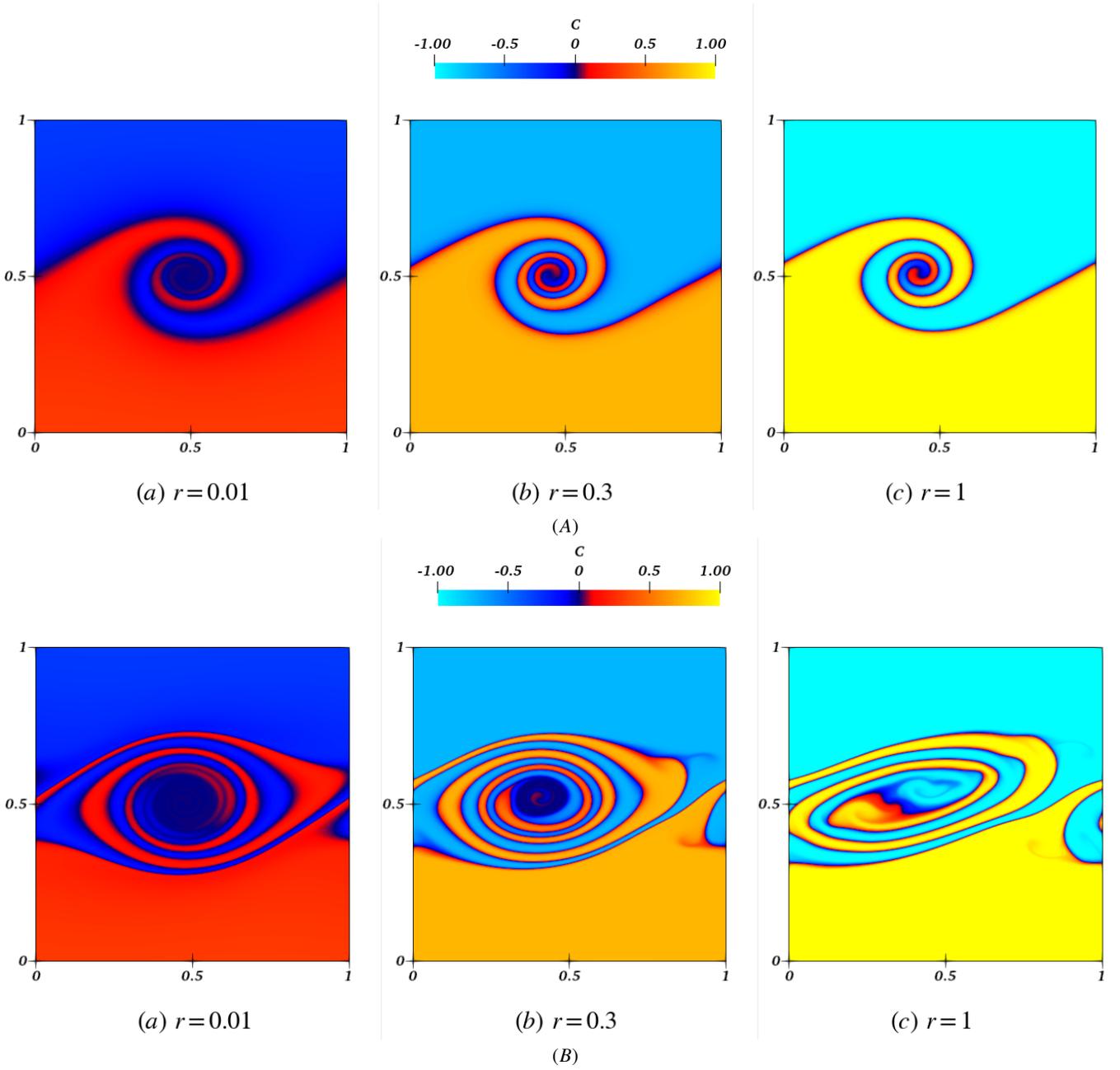

FIG. 14. Interface topology at $(A)$ $t = 2$ and $(B)$ $t = 8$ at smaller velocity profile thickness $\delta_u = 0.001$. The rest of the parameters pertinent to numerical simulations are $\rho_r = 0.8$, $Re = 10000$, $We = 1000$ and $Fr = 1$.

### 3. Effect of stratification

The magnitude of the stratification also plays a significant role in determining the propensity of the system towards the manifestation of the KH instability as it is the primary stabilizing agent. Since, the simulations in the current study are performed for $\rho_r = 0.8$, the extent of stratification can be controlled through the magnitude of the external acceleration controlled through the Froude ($Fr$) number. $Fr = 0.32$ is chosen for the analysis of higher stratification. The rest of the parameters pertinent to numerical simulations are $\rho_r = 0.8$,

$Re = 1000$, $We = 1000$ and $\delta_u = 0.01$. Fig. 16 presents the interface topology at $t = 8$ for the three values of the miscibility parameter. When the temperature of the system is far from the UCST ($r = 1$), the higher stratification leads to stable interface. The initially perturbed interface oscillates and eventually reaches a steady state horizontal configuration. Thus as evident from fig. 17, both the normalized potential energy and the normalized surface energy are constant throughout the process for the case $r = 1$. The kinetic energy however is lost due to diffusion. Further, the variance also remains constant. As the degree of partial miscibility increases, the density disparity between the two fluid layer falls leading to a weaker stratification. Consequently, both $r = 0.01$ and $0.3$ exhibit instability. While $r = 0.01$ again leads to the formation of billows like the case of $Fr = 1$, the case $r = 0.3$ does not allow the formation of vortex. The potential energy again remains constant. However, a marginal increase in the normalized surface energy is observed due to the formation of a sharp finger protrusion. Consequently, $r = 0.3$ leads to higher fall in kinetic energy as compared to $r = 1$. As the length of this protruded section increases, the variance in the domain also increases. Interestingly, the variance was found to increase almost linearly in this case (see fig. 17(d)). Finally, $r = 0.01$ exhibit high qualitative similarity for both the values of the $Fr$. Thus, one may conclude that for a fluid pair considered in equilibrium near the consolute point, i.e. for low values of $r$, the KH instability is qualitatively independent of the Froude number.

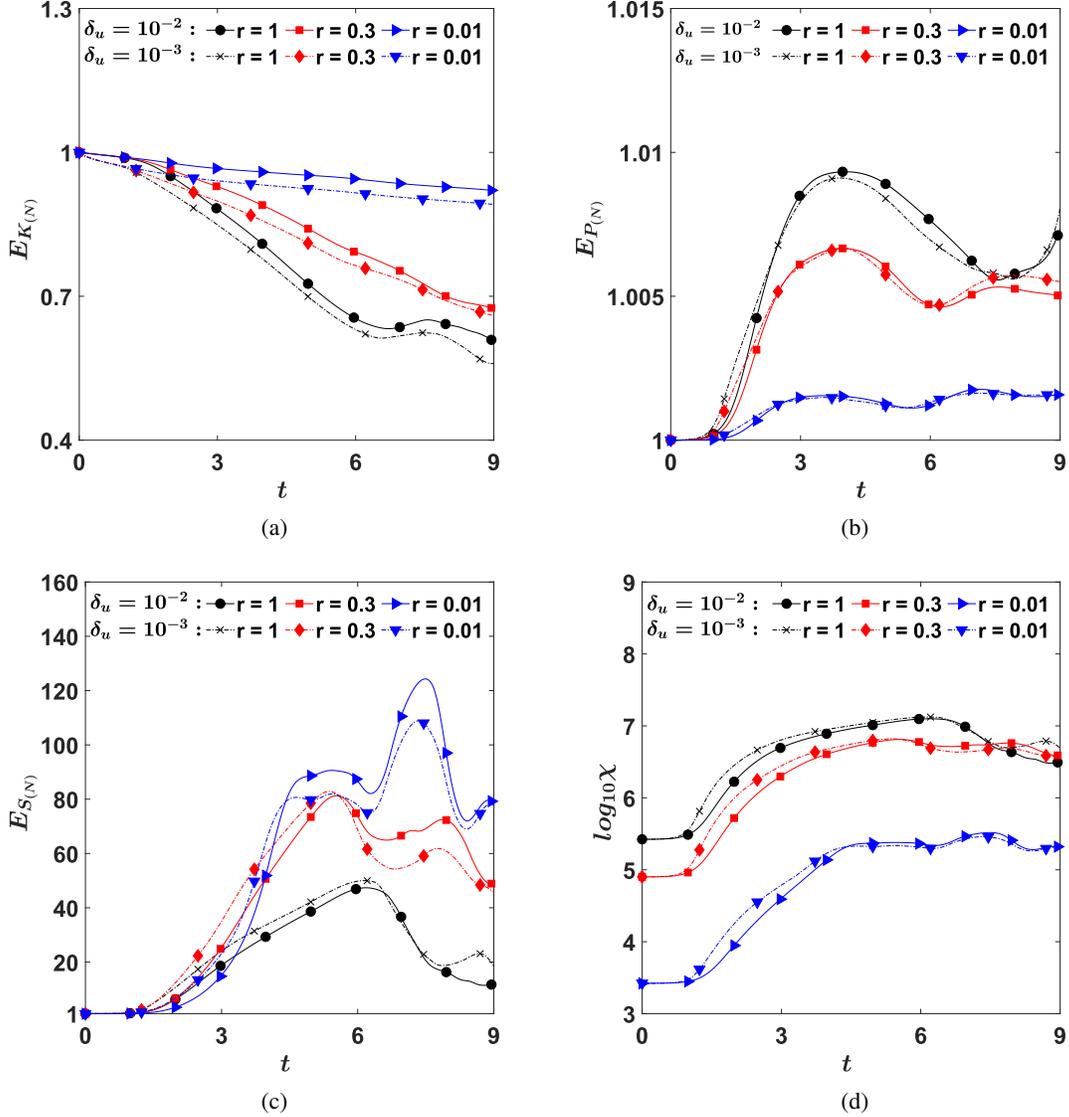

FIG. 15. Quantitative comparison of the effect of velocity profile thickness and degree of miscibility: (a) Normalized kinetic energy, (b) normalized potential energy, (c) normalized surface energy and (d) Variance of the density field. The rest of the parameters pertinent to numerical simulations are $\rho_r = 0.8$, $Re = 10000$, $We = 1000$ and $Fr = 1$.

### B. Initialization with thermodynamic non-equilibrium

In this section, we present our results for case $II$ pertaining to initialization with thermodynamic non-equilibrium. The instantaneous heating of the pair from a temperature below the UCST to a temperature above the UCST leads to simultaneous hydrodynamic flow as well as bilateral interfacial diffusion. Initially at $t = 0$, the two fluids are allowed to attain thermodynamic equilibrium at a temperature lower than the UCST characterized by $r = 0.01, 0.3$ and $1$. The system of fluids is then heated to a temperature above the UCST such that $r = -100$ in all the cases. It is driven from the initial state of immiscibility (or partial miscibility) towards the state of complete miscibility by the transformation of the double-well bulk free energy function to single well bulk free energy function. Typically in context of binary fluids with miscibility gap, the rate of diffusion is much smaller as compared to the development of hydrodynamic instability. In our previous study[41], we analyzed the system response in context of Rayleigh-Taylor instability for $r = -0.5$, i.e., heated to a temperature just above the UCST. A large difference between the time scale of the instability and interfacial mass transfer was observed thereby revealing very marginal differences between equilibrium and non-equilibrium configurations. Thus in the current work, the fluid pair is assumed to be heated to a temperature much higher than the UCST but below the boiling point of individual fluid layers therefore ensuring that the final value of $r$ is chosen in a way to obtain high enough rate of diffusion to reveal the competitive behavior in such configurations. One may also see eq. 6 to understand the role played by the value of the miscibility parameter $r$ in ensuring dominance of diffusion terms over compression term (the laplacian term). A qualitatively similar behavior is expected for all $r < 0$. In our previous study[41], we found the Weber number at the immiscible limit (or the surface tension $\tilde{\sigma}_0$) to play an important role in segregating the equilibrium cases from the non-equilibrium cases. Thus, two different values of the Weber number ($We = 100$ and 1000) are used for the current analysis. The rest of the parameters pertinent to numerical simulations are $\rho_r = 0.8$, $Re = 1000$, $Fr = 1$ and $\delta_u = 0.01$.

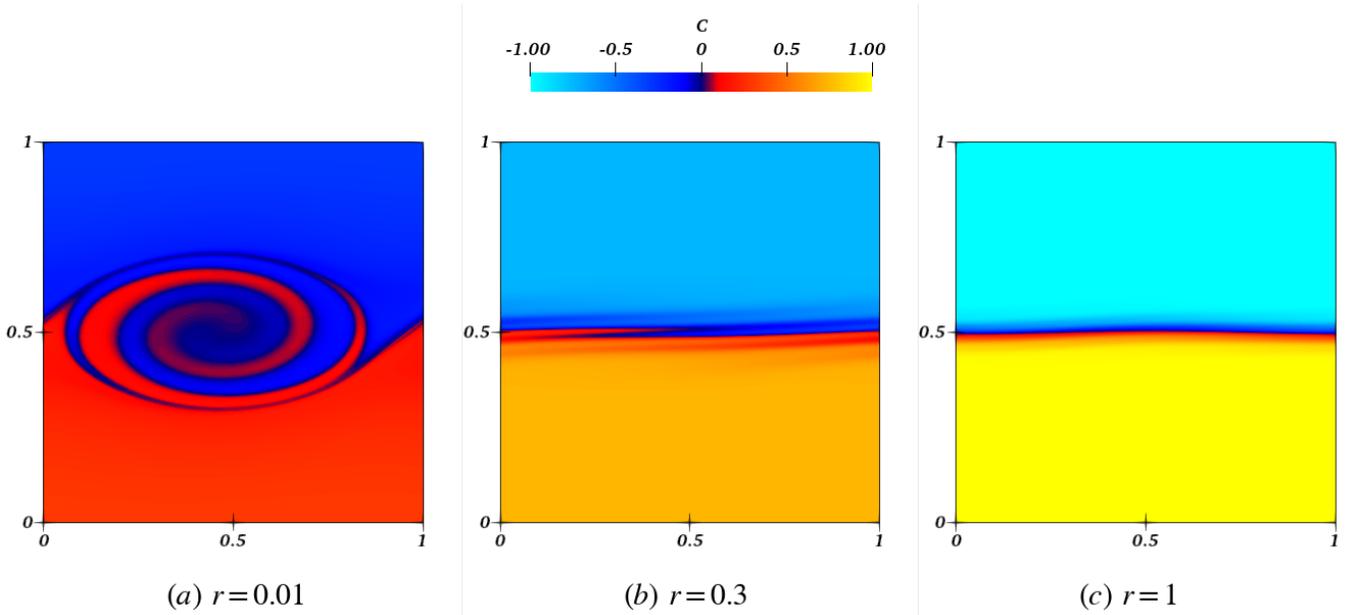

FIG. 16. Interface topology at $t = 8$ and $Fr = 0.32$ for (a) $r = 0.01$, (b) $r = 0.3$ and (c) $r = 1$. The rest of the parameters pertinent to numerical simulations are $\rho_r = 0.8$, $Re = 1000$, $We = 1000$ and $\delta_u = 0.01$.

Let us again start the analysis with $We = 100$. In the previous section, we elucidated the overall flow behavior in the context of equilibrium configuration (also see fig. 10 and fig. 13). Herein, we present the interface topology for the non-equilibrium configuration in fig. 18. The quantitative comparison between the two configurations is depicted in fig. 19. The mass transport across the interface results in significant qualitative as well as quantitative differences between the two cases. Again, one needs to look at the figs. 18 and fig. 19 in unison to understand the flow behavior. Due to the diffusion across the interface, the rate of mixing is enhanced. While the initial condition in both equilibrium and non-equilibrium configuration is same, the variance of the density field is governed by the competing effects of advection induced interface sharpening and the interfacial mass diffusion. The interfacial mass diffusion is governed by the gradient of the chemical potential, which inherently involves the $We$ number and $r$. Interestingly, for $r = 1$ and 0.3, the diffusion dominates at early time stages and therefore the variance decreases (see non-equilibrium case in fig. 19(d)). This period is followed by a period of domination of the advection induced interface sharpening leading to rise in the variance. Finally at later time stages, the diffusion again dominates leading to fall in variance. However, on the other hand, for $r = 0.01$, the gradient in chemical potential is not strong enough for the diffusion to dominate over advection induced sharpening at early time stages. Therefore, a marginal increase in the variance is observed. The rest of the two stages are same as that observed for $r = 0.3$ and 1. A significant difference is obtained between the variance at $t = 8$ for $r = 0.01$ between equilibrium and non-equilibrium cases. A comparison of the interface topology in fig. 10(a) and fig. 18(a) corroborates this observation. Further, the decrease in variance also explains the trend followed by the normalized surface energy. In the equilibrium configuration, the advection induced sharpening coupled with the deformation of the interface lead to a rise in the surface energy within the system. The non-equilibrium configuration, on the other hand, involves eventually decreasing surface energy of the system. Whereas, the normalized potential energy exhibits higher rise for non-equilibrium configuration as compared to the equilibrium configuration irrespective of the initial degree of miscibility between the two fluids as shown in fig. 19(b). The interfacial mass diffusion allows pockets of heavier fluid from the bottom layer to cross-over the interface towards the lighter fluid and vice-versa. Thus, the re-organized density field coupled with the elevation gained by the heavier fluid and the elevation lost by the lighter fluid explains the rise in the potential energy. The base shearing flow in each layer of the fluid then subsequently deforms the order parameter profile as evident from the comparison between fig. 10(b,c) and fig. 18(b,c). Finally, the normalized kinetic energy variation shown in fig. 19(a) can be explained by coupling the behavior of the surface energy and the potential energy of the system. As previously mentioned, the initial kinetic energy is the driving source and the other two components rise at the expense of the kinetic energy.

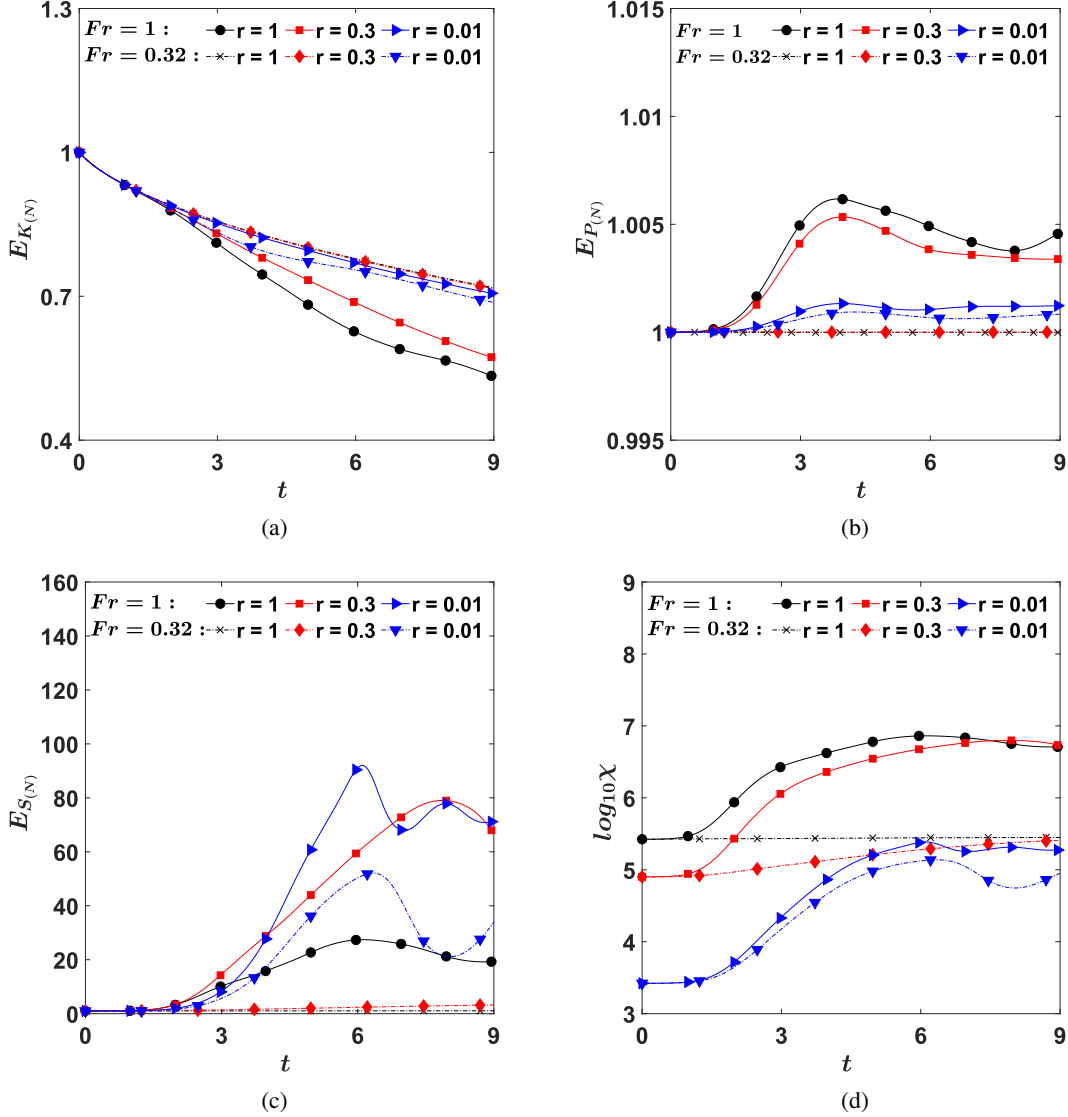

FIG. 17. Quantitative comparison of the effect of stratification and the degree of miscibility: (a) Normalized kinetic energy, (b) normalized potential energy, (c) normalized surface energy and (d) Variance of the density field. The rest of the parameters pertinent to numerical simulations are $\rho_r = 0.8$, $Re = 1000$, $We = 1000$ and $\delta_u = 0.01$.

Similarly, the flow behavior can be explained for the case of $We = 1000$. Figures 20 and 21 encapsulate the qualitative as well as quantitative aspects for this case. A qualitative comparison with equilibrium configuration can also be seen through fig. 11 which depicts the interface topology at $t = 8$. It is to be noted that only marginal differences are found while qualitatively comparing the equilibrium and non-equilibrium configuration. The higher value of $We$ in this case leads to relatively lower gradient in chemical potential and thus the system would require much longer time duration for significant differences to appear in the two configurations. The normal-ized kinetic energy and the potential energy shown in fig. 21(a,b) further corroborate the above-mentioned observation. We indeed observed quantitative differences in the normalized surface energy and the variance of the density field. Note that unlike the $We = 100$ case, we do not observe the initially diffusion dominated regime in temporal evolution of the variance even for $r = 1$ or $0.3$ (see fig. 21(d)). The normalized surface energy variation can be explained in a straightforward manner following the same arguments as the previous case.

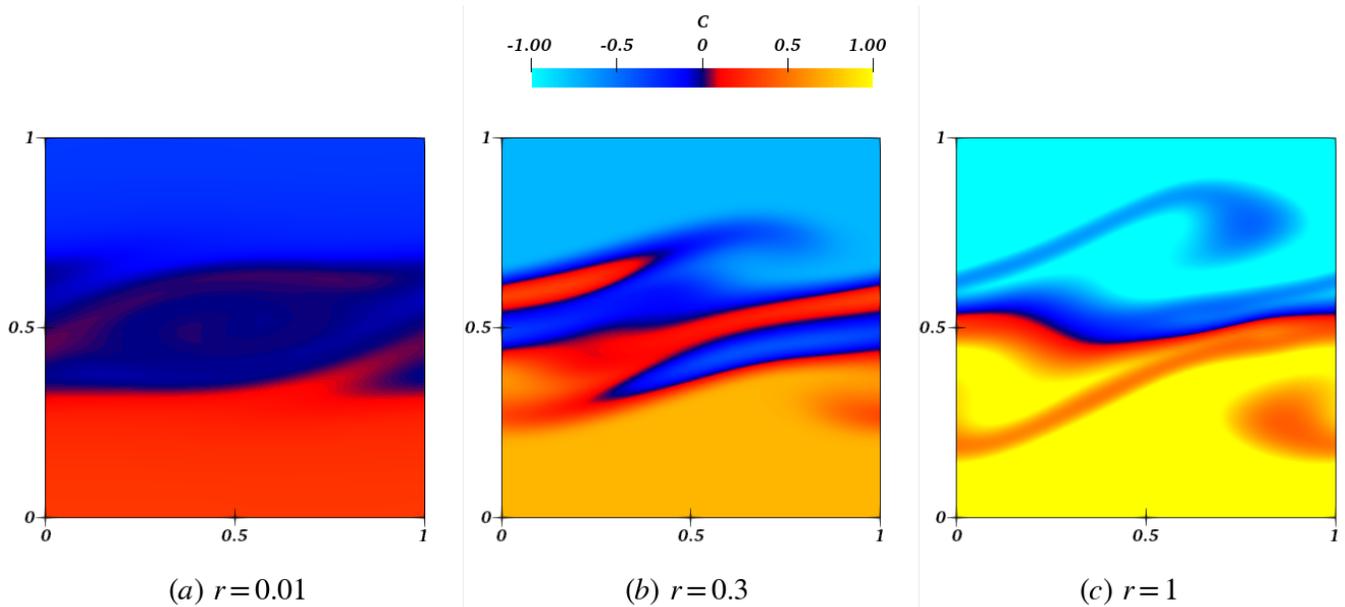

(a) $r = 0.01$    (b) $r = 0.3$    (c) $r = 1$

FIG. 18. Non equilibrium configuration: Interface topology at $t = 8$ and $We = 100$ for (a) $r = 0.01$, (b) $r = 0.3$ and (c) $r = 1$. The value of the miscibility parameter is switched to $r = -100$ at $t = 0$ right after initialization. The rest of the parameters pertinent to numerical simulations are $\rho_r = 0.8$, $Re = 1000$, $Fr = 1$ and $\delta_u = 0.01$.

Therefore, for the non-equilibrium configuration, the $We$ i.e. the surface tension at the limit of immiscibility and the final value of the miscibility parameter $r$ i.e. the temperature above UCST to which the fluid pair is heated plays the key role in determining the flow pattern. Naturally, the initial state of the system i.e. the initial degree of miscibility further governs the gradient in chemical potential and thus the three parameters should be studied in unison for any predictive analysis.

## VI. CONCLUSION

In this study, the isothermal evolution of an interface between binary fluids subjected to shear flow is investigated. A modified phase-field approach that incorporates the miscibility effects through a tuneable framework is employed to reveal the flow behavior in a broad parametric space. The degree of miscibility is governed by a dimensionless parameter $r$ which quantifies the proximity of the system temperature to the UCST of the fluid pair. The system of fluids subjected to shear flow is investigated for both equilibrium and non-equilibrium configurations at varying degree of partial miscibility.

For the first configuration where the two fluids are allowed to attain a state of thermodynamic equilibrium based on the system temperature, the effects of Weber number, velocity profile and the degree of stratification were analyzed. A thorough energy budget analysis entailing the kinetic energy, potential energy and the surface energy is performed to reveal the temporal evolution of the instability. While a monotonic trend for kinetic and potential energies is observed at higher Weber number ($We = 1000, 10000$), the lower $We = 100$ case results in interesting retraction of the interface. A further nuanced effect of mixing process is analyzed by computing the variance of the density field which quantifies the degree of non-homogeneity in the domain. Our numerical simulations accurately captured the advection induced interface sharpening and the corresponding effect on the surface energy of the system. Subsequently, we exploit the independent treatment of the velocity profile thickness and the order parameter profile thickness to investigate the effect of sharpening of the velocity profile. Since the velocity profile is sharper ($\delta_u = \frac{1}{10}\delta_c$) than the thinnest interface configuration ($r = 1$), the results indicate qualitatively independent growth of the instability irrespective of the proximity of the system temperature to the UCST as compared to the case when $\delta_u = \delta_c$. Finally, the combined effect of miscibility and the degree of stratification is analyzed by varying the Froude number ($Fr$). Interestingly, the reduced density disparity near the UCST led the system to be independent of the $Fr$.

The second configuration on the other hand pertains to non-equilibrium cases. The fluid pair while initialized at varying degree of partial miscibility is instantaneously heated to a temperature above the UCST ($r < 0$) to investigate the simultaneous effect of hydrodynamic instability and mass transport across the interface. Since the gradient of the chemical po-tential is the driving agent for the mass transport, the system exhibits dependence on the values of the $We$ and $r$. The interface evolution is governed by the competition between the advection based distortion and sharpening of the interface and the diffusion controlled growth of the interface. Our study reveals that based on the $We$ and the initial degree of partial miscibility, one may get three distinct growth regimes. For low $We$ and high initial $r$, the early times stages are marked with diffusion dominated regime, followed by advection induced sharpening dominated regime and finally the diffusion becomes prominent again. However, if the initial $r$ is very low or the $We$ is high, the first diffusion dominated regime is absent and the system of fluids begin the transition with advection dominated regime.

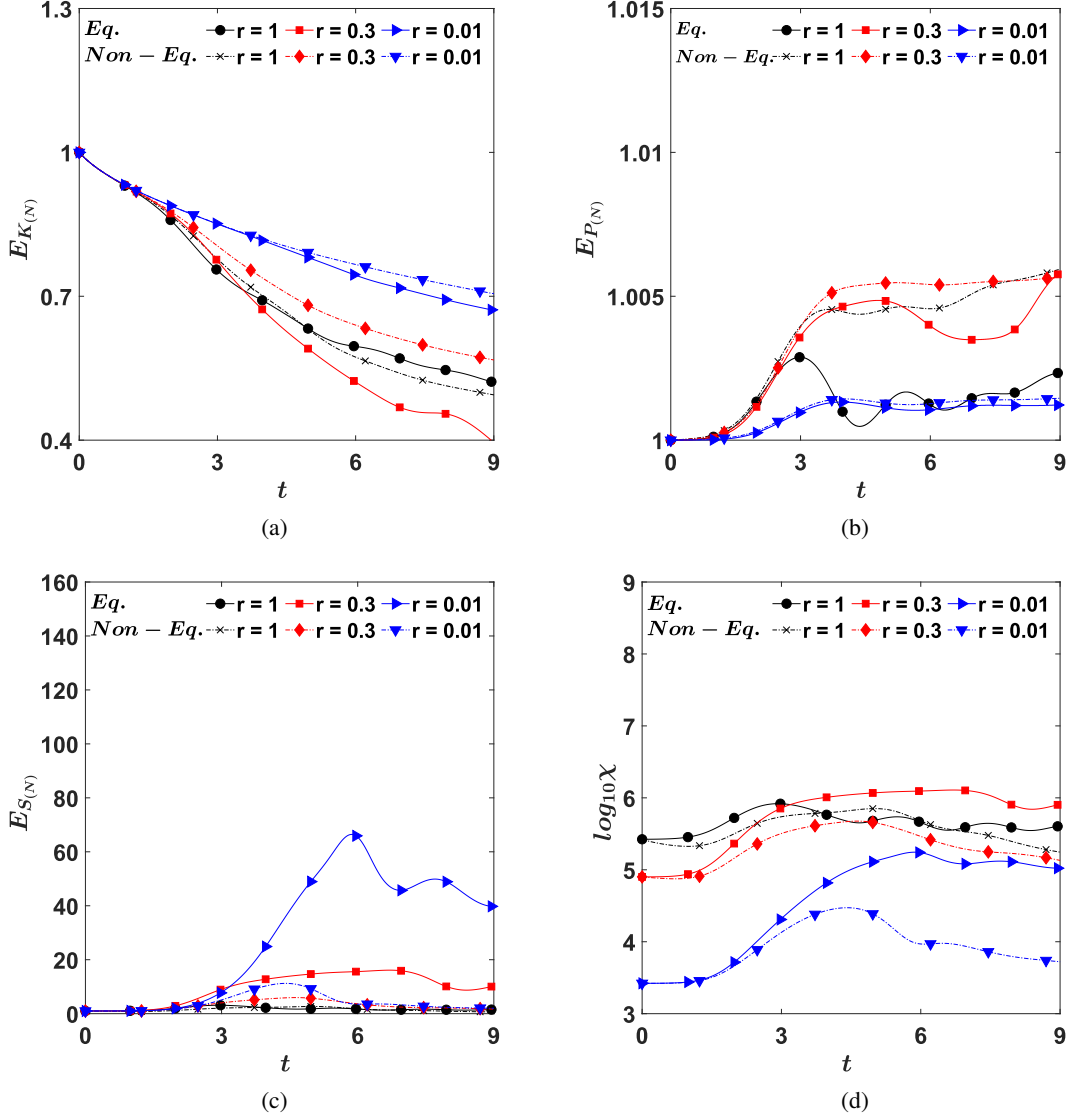

FIG. 19. Non equilibrium configuration: Quantitative comparison of the effect of mass transport across the interface at varying degree of initial miscibility at $We = 100$: (a) $E_{K_{(N)}}$, (b) $E_{P_N}$, (c) $E_{S_N}$ and (d) $\chi$. The rest of the parameters pertinent to numerical simulations are $\rho_r = 0.8$, $Re = 1000$, $Fr = 1$ and $\delta_u = 0.01$.

This study however deals purely with isothermal evolution of the instability. The incorporation of non-isothermal effects, which may also lead to coupled Marangoni flow is an interesting avenue for further research. Also, a non-isothermal analysis would require the space-dependent treatment of the miscibility parameter $r$ that may allow localized diffusion or phase-segregation. This involved and complex flow dynamics shall form the subject matter for our subsequent studies.

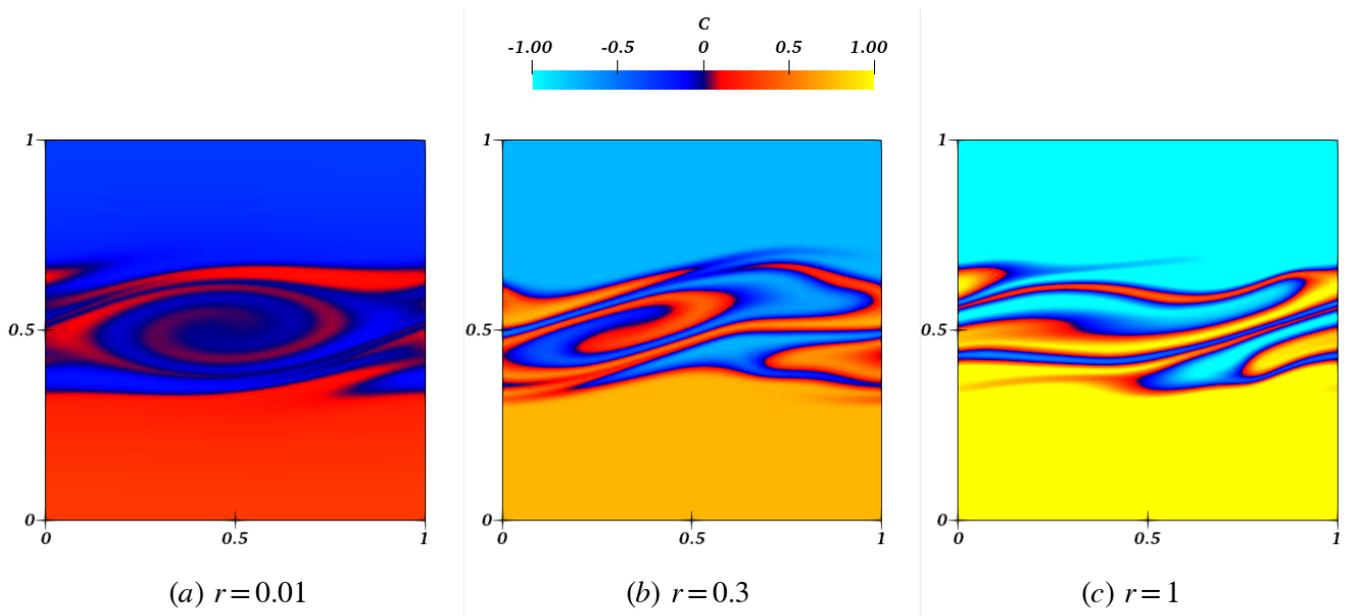

*(a)* $r = 0.01$        *(b)* $r = 0.3$        *(c)* $r = 1$

FIG. 20. Non equilibrium configuration: Interface topology at $t = 8$ and $We = 1000$ for (a) $r = 0.01$, (b) $r = 0.3$ and (c) $r = 1$. The value of the miscibility parameter is switched to $r = -100$ at $t = 0$ right after initialization. The rest of the parameters pertinent to numerical simulations are $\rho_r = 0.8$, $Re = 1000$, $Fr = 1$ and $\delta_u = 0.01$.

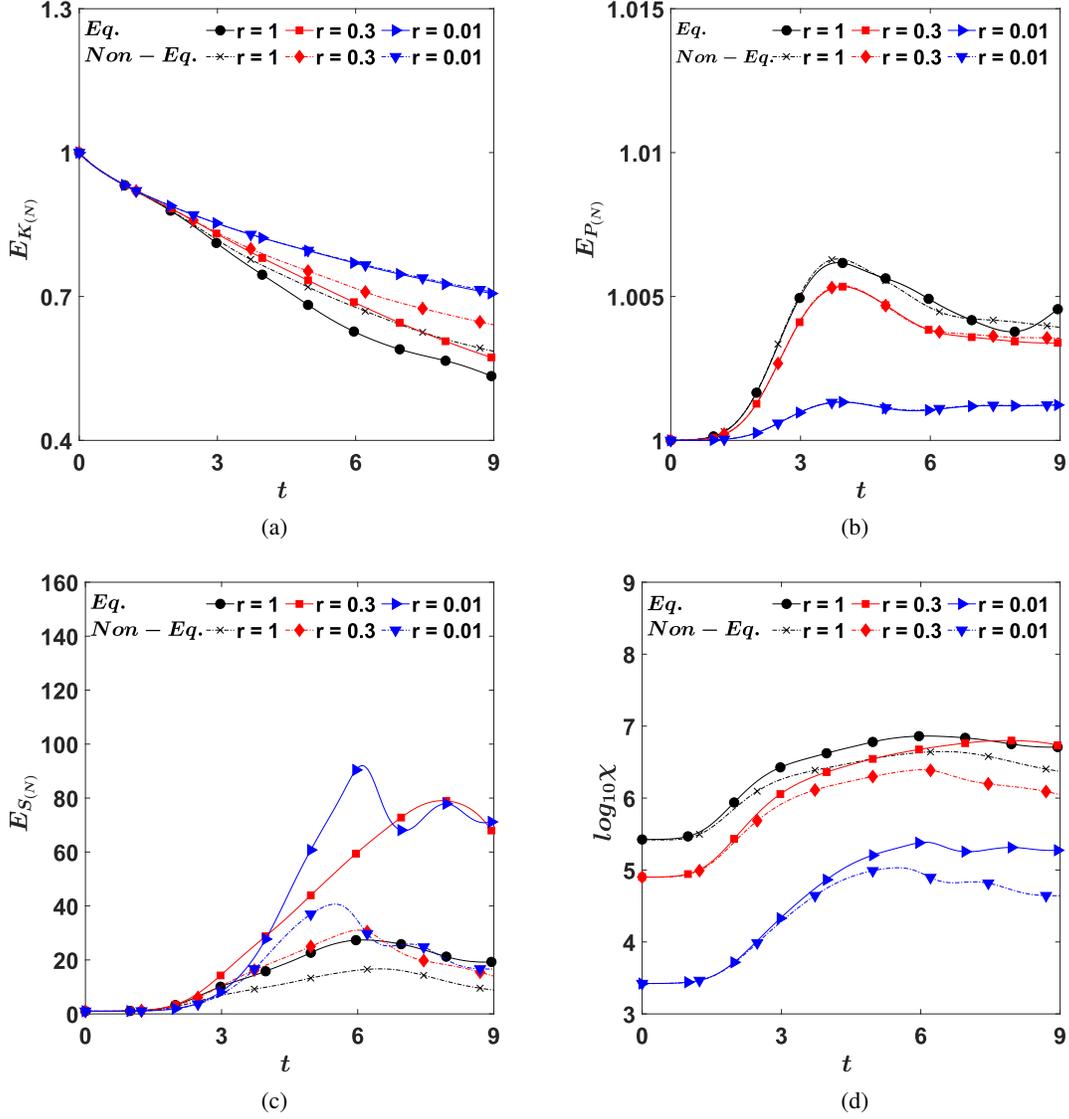

FIG. 21. Non equilibrium configuration: Quantitative comparison of the effect of mass transport across the interface at varying degree of initial miscibility at $We = 1000$: (a) $E_{K_N}$, (b) $E_{P_N}$, (c) $E_{S_N}$ and (d) $\chi$. The rest of the parameters pertinent to numerical simulations are $\rho_r = 0.8$, $Re = 1000$, $Fr = 1$ and $\delta_u = 0.01$.

## ACKNOWLEDGMENTS

The authors gratefully acknowledge the computing time provided to them on the high-performance computer Lichtenberg II. This is funded by the German Federal Ministry of Education and Research (BMBF) and the State of Hesse.

## FUNDING

A.D. and S.A. gratefully acknowledge the support provided by CNES (Centre national d'études spatiales: grant 9384-4500082826).

## CONFLICT OF INTEREST

The authors have no conflict of interest to disclose.

## DATA AVAILABILITY

The data that support the findings of this study are available from the corresponding author upon request.